\documentclass[]{aastex62}
\pdfoutput=1
\usepackage{natbib}
\usepackage{graphicx}
\usepackage{amsmath}
\usepackage{amssymb}
\usepackage{ctable}
\begin{document}
\title{Fitting the density substructure of the stellar halo with MilkyWay@home}
\author{Jake Weiss}
\affiliation{Rensselaer Polytechnic Institute}
\author{Heidi Jo Newberg}
\affiliation{Rensselaer Polytechnic Institute}
\author{Matthew Newby}
\affiliation{Temple University}
\author{Travis Desell}
\affiliation{University of North Dakota}

\begin{abstract}
We propose and test a method for applying statistical photometric parallax to main sequence turn off (MSTO) stars in the Sloan Digital Sky Survey (SDSS). Using simulated data, we show  that if our density model is similar to the actual density distribution of our data, we can reliably determine the density model parameters of three major substructures in the Milky Way halo using the computational resources available on MilkyWay@home (a twenty parameter fit).  We fit the stellar density in SDSS stripe 19 with a smooth stellar spheroid component and three major streams.  One of these streams is consistent with the Sagittarius tidal stream at $21.1$ kpc away, one is consistent with the trailing tail of the Sagittarius tidal stream in the north Galactic cap at $48$ kpc away, and one is possibly part of the Virgo Overdensity at $6$ kpc away.  We find the one sigma widths of these three streams to be $1.0$ kpc, $17.6$ kpc, and $6.1$ kpc, respectively.  The width of the trailing tail is extremely wide ($41$ kpc full width at half maximum).  This large width could have implications for the shape of the Milky Way dark matter halo.  The width of the Virgo Overdensity-like structure is consistent with what we might expect for a ``cloud"-like structure; analysis of additional stripes of data are needed to outline the full extent of this structure and confirm its association with the Virgo Overdensity.
\end{abstract}
\keywords{catalogs $-$ Galaxy: halo $-$ Galaxy: structure $-$ methods: data analysis $-$ methods: statistical}

\section{Introduction}

\subsection{Milky Way halo substructure}

The distribution of stars in the Galactic halo is dominated by dwarf galaxies and tidal streams of stars that have been stripped from them [see Fig. 1 of \cite{newberg2002}, and the ``Field of Streams" from \citep{Belokurov2006}].  These substructures represent the recent minor merger history of the Milky Way \citep{BullockJohnston2005}, and contribute to the build-up of stars in the Milky Way stellar halo.  Two or three dozen tidal debris streams, most of which extend tens of degrees or more across the sky, have been identified; the exact number cannot be determined due to controversy over the identity of tidal streams, particularly those discovered near the Galactic plane, and because some streams are detected at low enough significance that they are considered stream ``candidates." In addition to streams, other substructures of ambiguous origin (most notably ``clouds") continue to be discovered in the Galactic spheroid.  

For a review of tidal streams and clouds, see \cite{GrillmairCarlin2016}.  A more recent list of streams and clouds included in the `GALSTREAMS' Python Package can be found in Table 4 of \cite{Mateu2018}.  More recent halo substructures are identified in \cite{LiBalbinotMondrik2016}, \cite{Sohn2016}, \cite{Grillmair2017b,Grillmair2017a}, \cite{Jethwa2017}, and \cite{Shipp2018}. In particular, \cite{Shipp2018} identify eleven new substructures in the Milky Way halo using data from the Dark Energy Survey \citep[DES;][]{DESCollaboration2005, DESCollaboration2016}.  

Identification and measurement of tidal debris in the Milky Way halo is useful for understanding structure formation and galaxy assembly, and it has the potential to constrain the density distribution of the Milky Way's stellar halo.  Methods for measuring the halo shape from tidal streams have been, and continue to be, developed \citep[e.g.][for a review]{Law2010, Koposov2013, Kupper2015, Bovy2016, DierickxLoeb2017a, Sanderson2017, JohnstonCarlberg2016}.  In addition, the distribution of dark subhalos can be measured by looking for stars ejected from tidal streams \citep{SiegalGaskinsValluri2008}, stream heating \citep{Johnston2002} or stream gaps \citep{Carlberg2012}. \cite{Pearson2017} show that streams can also be used to constrain the rotation rate of the Galactic bar.
 
These techniques to determine the dark matter distribution in the Milky Way from tidal streams rely on accurate measurements of the tidal debris itself, but as we discover that halo tidal streams are more numerous and complex than originally thought, the association of particular stars with particular tidal streams becomes more ambiguous.  For example, \cite{Newberg2009} discovered that the blue horizontal branch stars (BHBs) thought to be associated with the southern portion of the Sagittarius (Sgr) dwarf tidal stream in \cite{Yanny2000} are actually part of the Cetus Polar Stream.  The Sgr dwarf tidal stream, which is the most prominent tidal stream in the sky, and the so-called ``bifurcated" Sgr stream that appears to split off from it, have also caused confusion; for example \cite{newby2013} suggested that the southern Sgr stream could be associated with the ``bifurcated" stream in the north, and the northern Sgr stream could be associated with the ``bifurcated" stream in the south.  These misidentifications and possible misidentifications of stars in the most prominent halo streams underscore the difficulties in counting and characterizing tidal streams.

\subsection{Statistical photometric parallax}

In this paper we present an improved {\it statistical photometric parallax} \citep{cole2008, newberg2013} method  to measure the spatial density of stars in the Milky Way stellar halo, using turnoff stars from the Sloan Digital Sky Survey \citep[SDSS;][]{SDSSYork}.  Statistical photometric parallax is the use of statistical knowledge of the distribution of the absolute magnitudes of stellar populations to determine the underlying density distributions of those stars.  This differs from photometric parallax in that the distance to each individual star is not determined.  

The idea of using turnoff stars to trace Milky Way halo substructure was introduced by \cite{newberg2002}.  They observed density substructure in the SDSS turnoff stars on the Celestial equator, and fit an absolute magnitude distribution to their blue turnoff star tracers.  In \cite{cole2008}, these tracers and a simplified absolute magnitude distribution from \cite{newberg2002} were used to build a model of the Milky Way halo and its substructure and complete preliminary fits to the stellar density of the halo and the Sagittarius dwarf galaxy tidal stream.  Several years later, \cite{newby2013} continued working with this model and showed that the massive distributed computing network, MilkyWay@home, could be effective in constraining the parameters in the density models of tidal streams.  Taking advantage of this new computational power, several optimizations were run on each stripe. Initially, the fitting algorithms were allowed great freedom in selecting the parameters.  Later, the parameters were constrained based on the results from neighboring stripes. 

These previous studies successfully used statistical photometric parallax to study the structure of the halo using SDSS turnoff stars.  SDSS turnoff stars, detected to a limiting magnitude of $g=22.5$, can be used to trace the structure of the Milky Way to 45 kpc from the Sun. However, the turnoff stars in a single stellar population, with the same color, can differ in absolute magnitude by two magnitudes (producing a distance error of a factor of 2.5).  Photometric parallax \citep[e.g.][]{Juric2008}  is unusable with turnoff stars because astronomers do not have a way to determine the distance to individual stars with reasonable accuracy using photometry alone.  

It has been shown that the absolute magnitude distribution of turnoff stars in halo globular clusters are surprisingly similar to each other, over a metallicity range -2.3$<$[Fe/H]$<$-1.2 dex and over ages ranging from 9 to 13.5 Gyr \citep{newby2011}. \cite{Grabowski2013} showed that this similarity holds even for the globular cluster Whiting 1, which is only 6 Gyrs old and has a metallicity of approximately [Fe/H]$\sim$0.6 dex \citep{Carraro2007, Valcheva2015}.  This surprising result, which comes about due to the age-metallicity relation for Milky Way stars, makes turnoff stars very useful for tracing the density of the stellar spheroid and outer disk.

In our work, we improve on the statistical photometric parallax methods by implementing a better model for the absolute magnitude of the tracer stars and their detection efficiency, and by using better fitting methods on MilkyWay@home.  In our implementation of statistical photometric parallax, we find the parameters in a density model that make the apparent magnitudes and angular positions of the observed stars most likely, using a maximum likelihood estimator (MLE) \citep{MLinAstronomy}.  The statistical description of the absolute magnitudes of the stellar tracers, and of the selection effects in the data, make statistical photometric parallax somewhat complex to apply. Because we are able to take all of these effects into account, we can reliably measure density distributions in real data.

There are four parts to statistical photometric parallax: data, a density model, an algorithm for measuring how well the model fits the data, and an algorithm for optimizing parameters.  The algorithm that measures how well the model fits the data includes the MSTO absolute magnitude distribution, as well as any observational biases.  It has taken us many years to perfect the algorithm that can simultaneously fit the spatial density of several tidal streams plus a smooth distribution to the SDSS MSTO stars; the smooth distribution represents the sum of: streams from small satellites, old streams that they are well mixed in density, and stars that were created during the collapse of the Milky Way, if any.

In this paper, we describe an improved algorithm for characterizing the spacial characteristics of stellar streams in the Milky Way halo using turnoff stars, and show that it is capable of simultaneously recovering the characteristics from three tidal streams plus a smooth halo component, using simulated data designed to mimic the stellar density in the actual Milky Way halo.  

\subsection{The big three halo substructures: the Sgr tidal stream, the ``bifurcated" stream, and the Virgo Overdensity}

\begin{figure}
\centering
\includegraphics[width=1.0\textwidth]{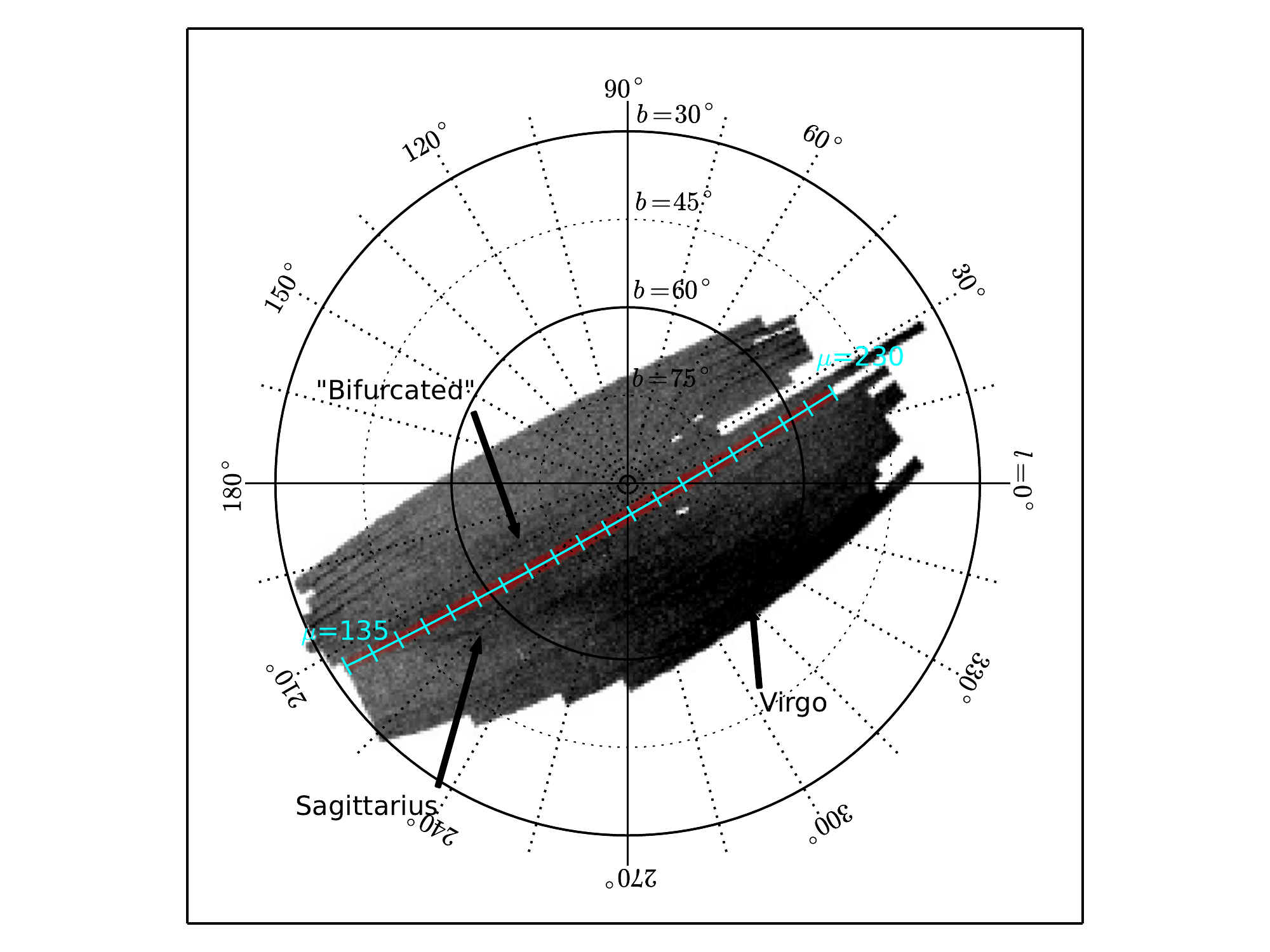}
\caption{The density of SDSS turnoff stars in the north Galactic cap. We highlight stripe 19 in magenta and show $\mu$ throughout this stripe in blue. There is some overlap between the stripes, which appears in this figure as dark streaks on the borders of the stripes.  The lowest latitude four stripes have a fainter magnitude limit, which explains why they appear darker.  In this plot, we can see the major substructures in our data, including the Sagittarius stream, the ``bifurcated" stream, and the Virgo Overdensity, with relation to the stripe we are fitting.  Note that stripe 19 passes through the Sgr dwarf tidal stream and the ``bifurcated" stream, but is far from the highest density region of the Virgo Overdensity.}
\label{SDSSNorth}
\end{figure}

We will present preliminary results for one $2.5^\circ$-wide SDSS stripe (stripe 19) of data that cuts across the northern Galactic hemisphere.  Figure \ref{SDSSNorth} shows the position of stripe 19 in the SDSS northern footprint.  The results from this stripe provide measurements of the largest known substructures in the Milky Way halo: the Sgr dwarf tidal stream, the so-called ``bifurcated" stream, and the Virgo Overdensity.  Stripe 19 crosses the Sgr dwarf tidal stream and the bifurcated stream, in a region of the sky in which they are clearly separated.  Since stripe 19 is more than 20 degrees from the densest portion of the Virgo Overdensity, it is uncertain whether a third substructure measured here is in the tails of the Virgo Overdensity, or whether it is associated with a new halo substructure.  In addition to three substructures, we fit smooth Milky Way halo and thick disk distributions.

The Sgr dwarf galaxy was first discovered by \cite{Ibata1995}, who found evidence of a dwarf galaxy within 16 kpc of the Galactic center, on the far side of the Milky Way, that was thought to be in the process of tidally disrupting.  The tidal stream of stars stripped from this dwarf galaxy have since been found to dominate the substructure of the Galactic halo \citep[e.g.][]{newberg2002,Majewski2003,Belokurov2006,Hernitschek2017}.  Though the Sgr dwarf galaxy and the stream of stars that have been tidally stripped from its gravitational grasp have been studied extensively \citep[see][for a recent review]{LawMajewski2016}; we are only starting to understand the dynamical history of this present-day merger.  

It has been a challenge to find a disruption model that simultaneously fits the positions of the leading and trailing tidal streams in the sky, the line-of-sight velocities of the stream stars, and the observed extension of the trailing tidal tail to $\sim$100 kpc from the Galactic center \citep{Newberg2003,Belokurov2014}.  \cite{DierickxLoeb2017b} present a recent simulation of the tidal debris that reproduces most of the measurements of the position of the leading and trailing tidal debris, including the distant stars in the trailing tidal tail and the observed ``spurs" at apogalacticon \citep{Sesar2017}, but still doesn't reproduce the line-of-sight velocities of the leading tail.  A previous model by \citet{Law2010} was able to fit the velocities of the leading tidal tail, using a triaxial dark halo model in which the disks rotate around the intermediate axis.  However, this Milky Way configuration is very unlikely \citep{Debattista2013}.  Refining the spatial distribution of the Sgr dwarf tidal stream using the algorithm described in this paper will help constrain N-body simulations of the Sgr dwarf tidal disruption, and lead to a better understanding of the shape of the Milky Way's dark matter halo.

The ``bifurcated" stream can be seen clearly in the ``Field of Streams" as a lower surface brightness companion stream to the Sgr dwarf tidal stream \citep{Belokurov2006}.  Belokurov identifies the Sgr dwarf tidal stream as ``Stream A," the ``bifurcated stream" as Stream B, and tentatively identifies a more distant ``Stream C" behind Stream A, which is now generally associated with an extension of the Sgr trailing tidal tail \citep{LiSmith2016}.  \citet{Koposov2012} shows the analogous bifurcated stream in the south Galactic cap.  Although the origin of the second, lower surface brightness stream close to the Sgr stream is not known, a leading possibility is that it could arise from multiple wraps of the stream around the Milky Way \citep{Fellhauer2006}.  Since its discovery, this stream has remained relatively unstudied compared to its sibling. \cite{Newberg2007} derive distances that are slightly farther away than Sgr for the bifurcated stream.  In contrast, \citet{NiedersteOstholt2010} says the bifurcated stream is slighly closer to the Sun than the Sgr dwarf tidal stream and \cite{Ruhland2011} finds the distances are basically the same.  \citet{Slater2013} show that the southern bifurcated stream is closer to the Sun than the southern portion of the Sgr dwarf tidal stream.  \citet{Yanny2009} show that the velocities and metallicities along the bifurcated stream are similar to those in Sgr. In \cite{Koposov2013}, there is evidence presented that the two streams may both pass through the progenitor, but due to the proximity of the progenitor to the Galactic bulge, it is difficult to see where exactly the two cross in reference to the progenitor.  In \cite{newby2013} it is suggested the streams may be from two separate progenitors that accreted around the same time, but the evidence to support this is not strong.  \citet{Hernitschek2017} give a possible fit to the bifurcated stream.  Currently, the origin of this stream is still an open question that our results will help answer.  Determining the origin of the ``bifurcated" stream is critically important, as it is useful for constraining the Milky Way potential \citep{Law2010, ViraCiro2013}.

The Virgo Overdensity/Virgo Stellar Stream \citep{Vivas2001,Juric2008,Duffau2006,Newberg2007} is a third large halo overdensity in the northern Galactic hemisphere, at distances of $6-20$ kpc from the Sun.  It is unclear whether this feature is a tidal stream, a ``cloud," or a combination of several different pieces.   \cite{CarlinVirgo} fit an orbit to the puffy structure, and suggest that this overdensity is the result of a recently disrupted massive ($10^9 M_\odot$) dwarf galaxy. \cite{CarlinVirgo} also finds their orbit includes the Pisces Overdensity. \citet{LiBalbinotMondrik2016} suggest Virgo could instead be associated with the Hercules-Aquila Cloud and Eridanus-Phoenix overdensities, since they are on the same polar plane, have similar galactocentric distances (18 kpc), and are separated by 120 degrees.  In \citet{Bonaca2012}, it is suggested that Virgo is ``cloud-like" and may have been the result of a minor merger that passed close to the Galactic center. \citet{Vivas2016} find several different, presumably unrelated, substructures of RR Lyrae stars at distances of 10-20 kpc in the Virgo region, and suggest there could be additional substructures at much larger distances.  The evidence for a more distant Virgo substructure is amplified by \citet{Sesar2017}, who find an outer Virgo overdensity at a distance of 80 kpc from the Sun.

The Milky Way stellar halo has traditionally been described by a smooth power-law distribution \citep[e.g.][]{Oort1975, Preston1991}.  Since the discovery of significant substructure in the stellar halo \citep{newberg2002}, researchers have had to choose whether to include or exclude these substructures when fitting the overall spheroid density. The smooth density component of the halo includes smaller or more thoroughly mixed remnants of tidal stripping, as well as any stars that were created in the initial gravitational collapse of the Milky Way galaxy. The algorithms used in this project will fit the smooth component and streams simultaneously.  Using simultaneous fitting for the background (as we will refer to the smooth component) and streams instead of subtracting a background to fit streams, we learn about the shape and density of this smooth component of the stellar halo without requiring a clean sky sample to fit it. 

\section{SDSS turnoff stars from stripe 19}\label{turnoffselection}

We will demonstrate this algorithm by fitting the density substructure of blue main sequence turnoff (MSTO) stars in SDSS stripe 19. Halo MSTO stars are more abundant than intrinsically brighter giant stars in the halo, and can be observed to distances of 46 kpc in SDSS photometric data.  Previous studies have found that MSTO stars bluer than the thick disk turnoff are present in the Sgr dwarf tidal stream, the ``bifurcated" stream, and the Virgo Overdensity.  Intrinsically fainter main sequence stars are not observed at distances far enough to trace our target halo substructures.

Whereas stars like red giant stars or BHBs are often assumed to have a known absolute magnitude based on their color or spectral properties, MSTO stars of a given color and stellar population are spread over a range of absolute magnitudes.  Instead of looking for a way to measure the absolute magnitude for each of these tracer stars, \cite{newberg2002} used the average apparent magnitude of the stars in a particular substructure, compared to the average absolute magnitude of the MSTO population, to ascertain the substructure's distance. \citet{cole2008} fit the density distribution of the Sgr dwarf tidal stream under the assumption that the absolute magnitude distribution of turnoff stars was Gaussian.  Later, by studying MSTO stars in Milky Way globular clusters, \cite{newby2011} not only fit a more accurate absolute magnitude distribution, which incorporated observational effects from SDSS and their selection efficiency, but also showed that the absolute magnitude distribution was the same in a range of globular clusters observed in the Milky Way halo.

We select our sample of turnoff stars from SDSS Data Release 7 \citep[DR7][]{DR7Abazajian2009}, with the criteria: $g_0 > 16$, $0.1<(g - r)_0<0.3$, $(u - g)_0 > 0.4$, and EDGE and SATURATED flags not set \citep{NewbergYanny2006, newberg2002}.  Selecting stars fainter than $g_0 = 16$ removes any saturated stars not identified with the saturated flag. The $(g - r)_0$ cut is used to pick out the blue side of the turnoff of the halo main sequence, while avoiding the redder thick disk turnoff stars. The $(u - g)_0 > .4$ cut is used to eliminate quasars. We use the subscript ``0" to indicate that the magnitudes we are using are reddening corrected using the \cite{SFD1998} dust maps.  To minimize disk contamination, stars with $b < 30^{\circ}$ are also cut from the data \citep{cole2008}.

The footprint of SDSS stripe 19 is shown in Figure \ref{SDSSNorth}. There are 84,046 turnoff stars (as selected by the cuts in the previous paragraph) in the magnitude range $16<g_0<22.5$, from SDSS stripe coordinates  $135^{\circ}<\mu<230^{\circ}$, and $-1.25^{\circ}<\nu<1.25^{\circ}$.

In some stripes, there are globular clusters or other stellar substructure such as the Monoceros ring \citep{newberg2002, Yanny2003} or the Galactic bulge that are not well fit by our parameterized density model.  We avoid these structures by removing the area of the sky in which they are contained.  We can remove a small area of the sky around globular clusters, and then also remove a section of the sky over which our model is integrated, as described in section \ref{SectionCut}.  Low latitude substructure can be removed by removing a larger area of data near the Galactic center and anti-center where necessary, thus making the stripe shorter.

An SDSS ``wedge" includes a volume defined by the angular limits of a stripe and the distance (from the Sun) to the most distant object in the dataset.  Each SDSS stripe is $2.5^\circ$ wide, and typically $140^{\circ}$ long.  Since the density varies only a small amount in the narrow direction, we often depict the stellar density in a polar plot with the radius proportional to the distance and the angle given by $\mu$, the angular distance along the stripe.  We apply our algorithm to one wedge (stripe) of data at a time.

\section{Parameterized Milky Way halo model}

In this section we describe the density model that we will fit to a single SDSS stripe.  This model is adapted from a model originally used in \cite{cole2008} and later used on the distributed computing platform MilkyWay@home by \cite{newby2013}. As color errors increase, MSTO stars are scattered outside the color selection bin and redder stars are scattered into the color selection bin.  This effect is especially pronounced near the survey detection limit where color errors are high. A major change we made to the model is the inclusion of these effects on our absolute magnitude distribution and completeness as described in \cite{newby2011}.  Although the color selection range was chosen to be bluer than the turnoff of the Milky Way thick disk, we have also added a thick disk component to the smooth portion of the density profile to take into account that a few of these stars might have leaked into our selection. 

Throughout this section we develop (for a single stripe) an MLE that measures how well a model with a particular set of parameters fits the data.  This estimator is then be used to optimize the model parameters. Although we will fit three streams plus the smooth component in this paper, the model as implemented can fit an arbitrary number of tidal streams or substructures in a given stripe.  The code for this release of our model can be found at \cite{MilkyWay}.

\subsection{Smooth component}

We implemented two different models for the smooth component of the halo, both with two tunable parameters.  One model is a Hernquist spheroid model \citep{hernquist, Xu2015} with a double exponential disk.  The other is a broken power law (BPL) \citep{shaila2012} without a disk.  In our model fitting, we will primarily use the Hernquist model in order to be consistent with previous versions of the model described in \citet{cole2008} and \citet{newby2013}. First we describe the Hernquist/thick disk model, and then the BPL model, which we used to study the effect of an imperfectly modeled smooth component on the derived properties of the halo substructure.

\subsubsection{Hernquist plus double exponential disk}

The Hernquist distribution \citep{hernquist} is described by the equation:
\begin{equation}
\rho_{spheroid}(r) \propto \frac{1}{r(r+r_0)^3}
\end{equation}
where $r = \sqrt{X^2 + Y^2 + \frac{Z^2}{q^2}}$; note that this is not spherical radius, but instead an ellipsoidal radius.  X, Y and Z are Galactocentric Cartesian coordinates with the Sun at $(-8.5, 0, 0)$ kpc, Y in the direction of the Sun's motion, and Z in the direction of the north Galactic pole. $r_0$ is a scale radius and $q$ is a flattening parameter.  Using this model, there are two tunable parameters: $r_0$ and $q$ .  After inspecting the Hernquist distribution \citep{hernquist}, the Einasto profile \citep{ einasto} and a broken power law distribution \citep{shaila2012}, we found that scale radius only holds a minor effect on the overall shape of the distribution over the range of our data. In our model, we fix the scale radius at 12.0 kpc, which approximately reproduces the average of all of the distributions inspected.

\begin{figure}
\centering
\gridline{
\includegraphics[width=0.49\textwidth]{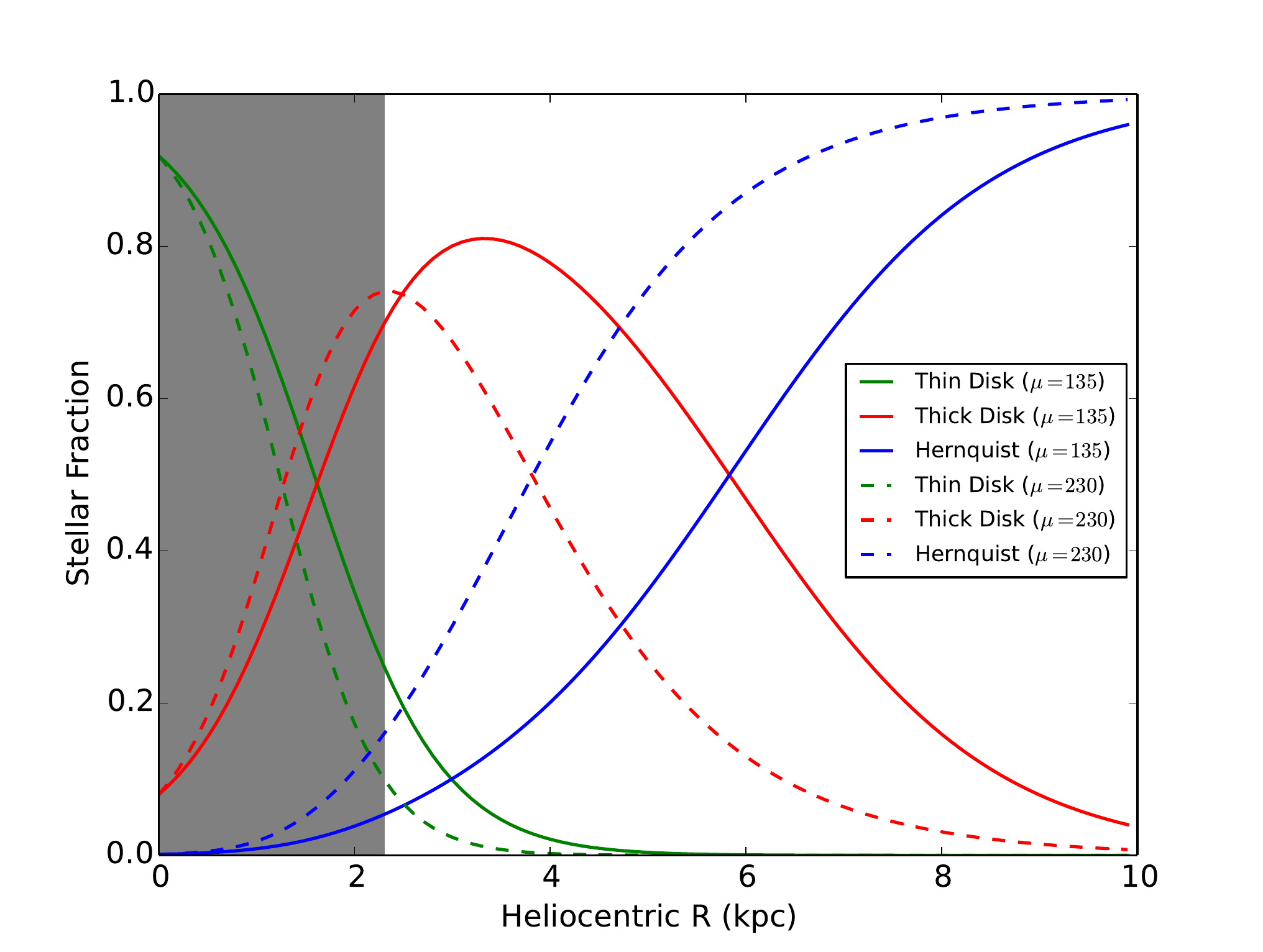}
\includegraphics[width=0.49\textwidth]{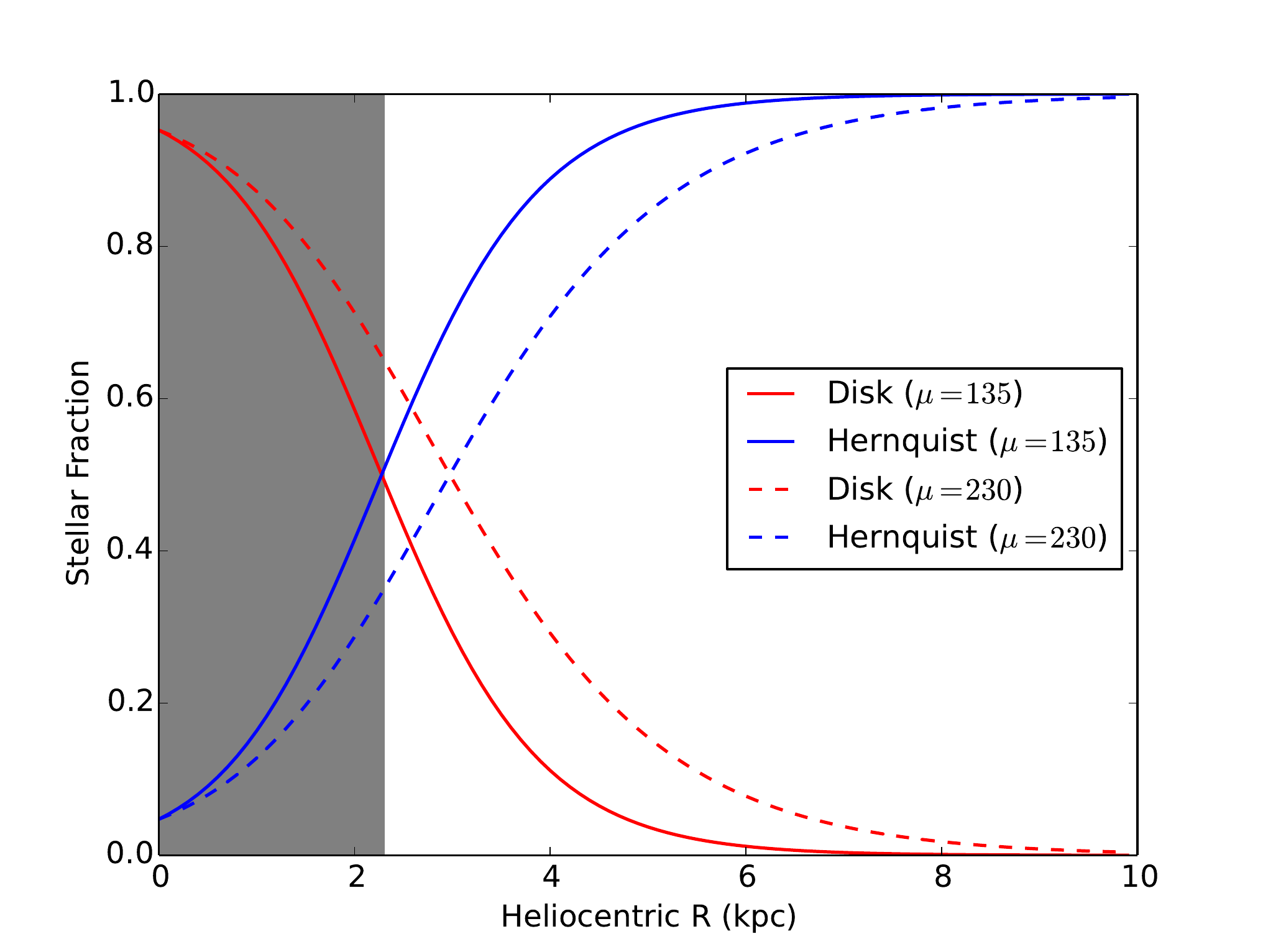}
}
\caption{Fraction of stars from the stellar halo, the thin disk and the thick disk as a function of heliocentric distance in the wedge.  The first panel shows the \cite{Xu2015} model and the panel on the right is our model after fitting stripe 19. The grey portion of the plot is outside of our data and everything to the right of it is inside of our data selection area.  The green lines represent the thin disk which while present in the \cite{Xu2015} model is not present in ours since it is negligible (and not fit) in the wedge. The blue line is the Hernquist background which should dominate in our data, and  the red line is the thick disk fraction which while very present in the \cite{Xu2015} model is reduced in our model since most of the disk stars are eliminated with our color cuts.}
\label{StarFraction}
\end{figure}

The double exponential thick disk used in our model is described by the equation:
\begin{equation}
\rho_{disk}(R_{cyl}, Z) \propto e^{R_{cyl}/l_{disk}} e^{|Z|/h_{disk}};
\end{equation}
where $R_{cyl} = \sqrt{X^2 + Y^2}$; $l_{disk}$ is the disk scale radius; and $h_{disk}$ is the disk scale height.  In our model, we fix the scale length and height of the disk to be the same as those found in \cite{Xu2015}: $l_{disk} = 3500$ pc and $h_{disk} = 700$ pc.  In Figure \ref{StarFraction}, we show the relative expected fractional contributions of the thick disk, thin disk and Hernquist background in the \cite{Xu2015} model at different positions in SDSS stripe 19.  We do not include the thin disk in our model because our data is far enough from the Galactic plane to avoid the majority of the stars in this component, and our color cuts also help eliminate disk contamination.

Because we are using blue MSTO stars that preferentially avoid the thick disk, we cannot use a published normalization for the disk and halo, so this normalization is a fit parameter. We write the overall density of the combined stellar spheroid and disk as:
\begin{equation}
\label{background}
\rho_{smooth}(X,Y,Z) = \epsilon_{spheroid} * \rho_{spheroid}(r) + (1 - \epsilon_{spheroid}) * \rho_{thick disk}(R_{cyl},Z),
\end{equation}
where $\epsilon_{spheroid}$ is a weighting factor that allows us to fit any mix of densities between all disk ($\epsilon_{spheroid}=0$) and all spheroid ($\epsilon_{spheroid}=1$).  The following equation can be used to convert $\epsilon_{spheroid}$ to $f_{disk}(X,Y,Z)$, the fraction of thick disk stars at a given point:
\begin{equation}
f_{disk}(X,Y,Z) = \frac{(1 - \epsilon_{spheroid}) * \rho_{thick disk}(R_{cyl},Z)}{\rho_{smooth}(X,Y,Z)}.
\end{equation}
To find the fraction of stars in the spheroid at a point, $f_{spheroid}(X,Y,Z)$, use:
\begin{equation}
f_{spheroid}(X,Y,Z) = 1 - f_{disk}(X,Y,Z).
\end{equation}
The two parameters that we fit in the smooth Hernquist/double exponential model are $q$ and $\epsilon_{spheroid}$.

\subsubsection{Broken power law model}\label{BPLSection}
The BPL distribution from \cite{shaila2012} is described by the following equation:
\begin{equation}
\rho_{BPL}(r) \propto (r/r_{0})^{n},
\end{equation}
where
\begin{equation}
n = 
\begin{cases}
-2.78 \text{ if } r < 45 \text{kpc} \\
-5.0 \text{ if } r \geq 45 \text{kpc}
\end{cases}\text{,}
\label{BPLPowers}
\end{equation}
$r_0 = 8.5$ kpc is the distance from the Galactic center to the Sun, and $r = \sqrt{X^2 + Y^2 + \frac{Z^2}{q^2}}$. We use  
\begin{equation}
q = 
\begin{cases}
q_0 + (1-q_0)(R_{cyl}/R_u) \text{ if } R_{cyl} < R_u \\
1.0 \text{ if } R_{cyl} > R_u
\end{cases}\text{,}
\label{BPLQ}
\end{equation}
where $q_0 = 0.5$, $R_{cyl} = \sqrt{X^2 + Y^2}$, and $R_u = 20$ kpc as written in \cite{Keller08}.  

We do not use this model for fitting the halo because the Hernquist model is more comparable to previous work.  Instead, we use this model for testing our algorithm.

\subsection{Stream}
In each stripe, which probes a wedge-shaped volume of the Galaxy, the density of a tidal stream or cloud is fit to a cylinder with a uniform density along its length, and Gaussian fall-off in density as a function of radius.  The orientation of the stream through the wedge is described by a unit vector $\hat{a}$ with a polar angle from the $Z$-axis of $\theta$, and an azimuthal angle of $\phi$ measured from the Galactocentric $X$-axis and increasing in the direction of the Galactocentric $Y$-axis.  The position of the cylinder's axis, at the point that it passes through the center of the stripe ($\nu=0$), is given by an angular coordinate along the stripe, $\mu$, and a distance from the Sun, $R$.  The $(\mu, \nu)$ coordinates are SDSS great circle coordinates \cite{SDSSYork}, which are defined separately for each stripe.

At point $\textbf{p}$ relative to the cylinder, the stream's density $\rho$ is described by the function: 
\begin{equation}
\rho_{stream}(\textbf{p}) \propto e^{\frac{-d^2}{2\sigma^2}}
\end{equation}
where $d$ is the radial distance from the stream, and $\sigma$ is the standard deviation describing the stream's width \citep{cole2008}. 

The final parameter in the stream model is the stream weight, $\varepsilon$, which measures the fraction of stars in the wedge that belong to a stream. To get the fraction of stars in a stream or smooth component (spheroid) from the weights we use the following functions: 
\begin{equation}
\begin{array}{l}
f_{stream_i}=\frac{e^{\varepsilon_i}}{1 + \sum^k_{j=1}[e^{\varepsilon_j}]} \\
f_{spheroid}=\frac{1}{1 + \sum^k_{j=1}[e^{\varepsilon_j}]}\text{.}
\end{array}
\label{weights}
\end{equation} 
Here, $i$ and $j$ denote the stream number for a total of $k$ streams.  The weight of a stream is used for model optimization instead of the total star counts because it is a continuous real number that is defined for all real numbers, making it a better parameter than star counts, which are discrete numbers \citep{cole2008}.

In total, six parameters are required to fit a single stream:  the 4 spatial coordinates, $\theta$, $\phi$, $\mu$, and $R$, which give the stream position and orientation; the width parameter, $\sigma$; and the stream weight, $\varepsilon$.  The model can handle an arbitrary number of streams in a single wedge.

\subsection{Absolute magnitudes of MSTO stars} \label{AbsoluteMagnitudesofMSTOStars}

In our implementation of statistical photometric parallax \citep{newberg2013}, we use the absolute magnitude distribution for MSTO stars described in \cite{newby2011}, which accounts for the distribution change due to contamination from redder stars at faint magnitudes.  In this implementation of statistical photometric parallax, we convolve the model density distributions (in the $R$ direction which corresponds to our line-of-sight) with the absolute magnitude distribution. The convolution is given by:
\begin{equation}
\rho^{con}_{comp}(l,b,\mathcal{R}(g_0)) = \frac{1}{\mathcal{R}^3(g_0)}\int^{\infty}_{-\infty}\mathcal{R}^3(g_0-g)\cdot\rho_{comp}(l,b,\mathcal{R}(g_0-g))\cdot N(g_0-g;g_0,u)dg
\end{equation}
where $\mathcal{R}(g_0)$ is distance to a star (or position in apparent magnitude space) with an apparent magnitude of $g_0$ and assumed absolute magnitude of $M_{g_0} = 4.2$, $\rho_{comp}$ is a density model, either background or stream, $l$ and $b$ are  Galactic $l$ and $b$, and $N(g_0-g;g_0, u)$ is our MSTO star absolute magnitude distribution.  This integral is the combination of the MSTO star absolute magnitude distribution function with our stellar density function over all magnitude space.  For an in depth explanation of this integral, see the original derivation in \cite{cole2008}.

In our model, we use the MSTO absolute magnitude distribution found in \cite{newby2011}, which is defined by two half Gaussian distributions which are normalized together:
\begin{equation}\label{TwoHalfGaussian}
N(x;g_0,u)=\frac{1}{\frac{(\sigma_l + \sigma_r(\mathcal{R}(g_0)))}{2}\sqrt{2\pi}}e^{\frac{-x^2}{2u^2}}
\end{equation}
where
\begin{equation}
u = 
\begin{cases}
\sigma_l =.36 \text{ if } x < 0 \\
\sigma_r(\mathcal{R}(g_0)) = \frac{\alpha}{1 + e^{-(\mathcal{R}(g_0) - \beta)}} + \gamma \text{ if } x \geq 0
\end{cases}\text{,}
\label{Modfit}
\end{equation}
$\alpha = 0.52$, $\beta = 12.0$ and $\gamma = 0.76$.  This absolute magnitude distribution accounts for an increase in the standard deviation on the faint side of the absolute magnitude distribution as a function of magnitude in the SDSS.  The change in distribution is due to the change in the type of stars found in our narrow MSTO color selection criteria as increasing color errors towards the magnitude limit of the survey. Toward the survey magnitude limit, color errors increase and cause redder, lower main sequence stars to leak into our color selection as actual MSTO stars leak out.  The underlying absolute magnitude distributions of all turnoff stars in the halo is assumed to be constant as found by \cite{newby2011}.

\subsection{Detection and selection efficiency}
To correct for the drop off in detection efficiency near the magnitude limit of the SDSS, a sigmoid curve was fit in \cite{colethesis} to data from \cite{newberg2002}:
\begin{equation}
\label{sig}
\mathcal{E}_{sigmoid}(g_0)=\frac{s_0}{e^{s_1(g_0-s_2)}+1},
\end{equation}
where $(s_0, s_1, s_2)=(0.9402, 1.6171, 23.5877)$. This curve describes the survey completeness at a given magnitude.  Using this curve, we can account for the decreasing completeness of the survey with increasing magnitude. 

\begin{figure}
\centering
\includegraphics[width=1.0\textwidth]{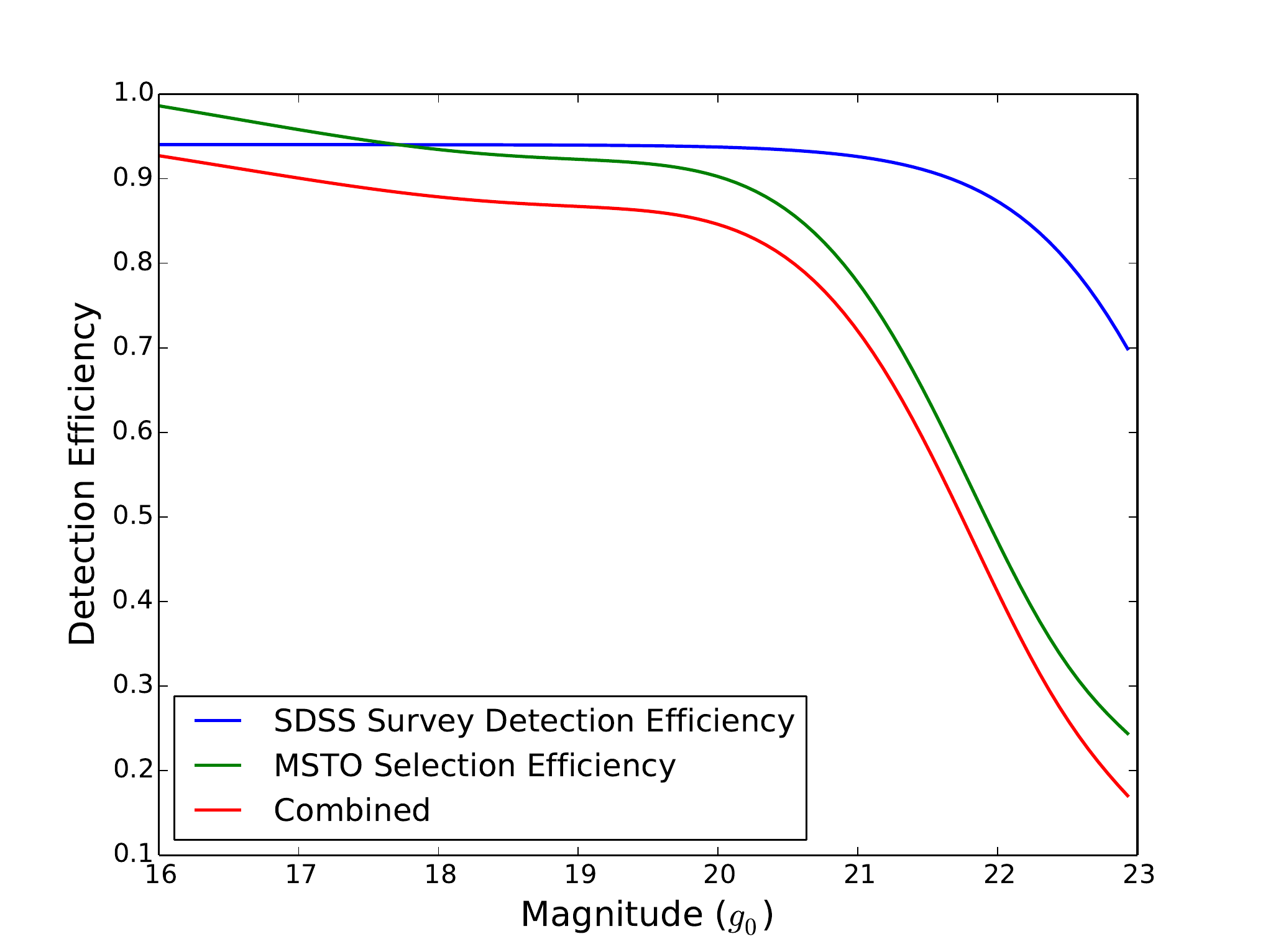}
\caption{Detection and selection efficiency as a function of magnitude.  The blue line shows the detection efficiency for the SDSS as fit by \cite{newberg2002}, the green line shows the detection efficiency correction to account for stars leaking in and out of the color selection bin \citep{newby2011}, and the red line shows the overall combined detection efficiency for MSTO stars in the SDSS survey.  The inclusion of the selection efficiency drastically changes the completeness for MSTO stars around 20th magnitude and dimmer.}
\label{ModfitDetection}
\end{figure}

As the color errors increase, fainter main sequence stars leak into the MSTO color selection box, and MSTO stars leak out of the MSTO color selection box.  The detection efficiency is adjusted using what we call the selection efficiency to account for the number of stars expected given this effect.  The selection efficiency for SDSS stars in the MSTO selection bin is given by a 7th order polynomial fit by \cite{newby2011}.  We reproduce it here because there was an inadvertent truncation of significant digits in the original paper that affects the result of the fit: 
\begin{equation}
\label{starleak}
\mathcal{E}_{selction}(\mathcal{R}(g_0))=\frac{n(\mathcal{R}(g_0))}{n_0}=\sum^7_{i=0}(a_{yi}+a_{ri})\mathcal{R}(g_0)^i
\end{equation}
where $a_y = (1.05628761, -3.14555041 \times 10^{-2}, 2.05499665 \times 10^{-4}, 2.53747387 \times 10^{-6}, -2.67000303 \times 10^{-8}, 0, 0, 0)$ and represents the remaining MSTO stars in the color selection bin and $a_r = (1.60879353 \times 10^{-2}, -1.97164570 \times 10^{-2}, -4.31844102 \times 10^{- 4}, 6.60960070 \times 10^{-3}, 1.26368065 \times 10^{-5}, -1.91560491 \times 10^{-7}, 1.47140445 \times 10^{-9}, -4.53857248 \times 10^{-12})$ and represents the red stars that leak into the color selection bin to contaminate the sample. The percentage of stars left in the selection bin assuming 100 percent detection efficiency is illustrated in Figure \ref{ModfitDetection}.  The two detection efficiency functions are multiplied together to get the total detection efficiency for stars that fit the MSTO selection criteria.

\subsection{Maximum likelihood estimator}

Putting together all of the individual model components, we create an MLE which tells us how well our model, with parameters $\textbf{Q}$, fits a set of data ($N$ stars with measured $l_i, b_i, g_i$):
\begin{equation}
\mathcal{L}(\textbf{Q}) = \prod^N_{i=1}PDF(l_i,b_i,R(g_i)|\textbf{Q}),
\end{equation} 
where $i$ is the index of each star, $l$ and $b$ are Galactic coordinates \citep{cole2008}. Substituting in our density functions, correcting for imperfect knowledge of the absolute magnitudes of the stars, and taking into account the detection efficiency, our PDF is given by:
\begin{multline} \label{PDFEqn}
PDF(l,b,\mathcal{R}(g_0)|\textbf{Q}) = \frac{1}{1+\sum^n_{i=1}e^{\epsilon_i}} \frac{\mathcal{E}(\mathcal{R}(g_0))\rho^{con}_{background}(l,b,\mathcal{R}(g_0)|\textbf{Q})}{\int\mathcal{E}(\mathcal{R}(g_0))\rho^{con}_{background}(l,b,\mathcal{R}(g_0)|\textbf{Q})dV} + \\ \sum^n_{i=1}\frac{e^{\epsilon_i}}{1+\sum^n_{j=1}e^{\epsilon_j}} \frac{\mathcal{E}(\mathcal{R}(g_0))\rho^{con}_{stream_i}(l,b,\mathcal{R}(g_0)|\textbf{Q})}{\int\mathcal{E}(\mathcal{R}(g_0))\rho^{con}_{stream_i}(l,b,\mathcal{R}(g_0)|\textbf{Q})dV},
\end{multline}
where $n$ is the number of streams in the model and $\mathcal{E}(\mathcal{R}(g_0)) = \mathcal{E}_{sigmoid}(\mathcal{R}(g_0)) \cdot \mathcal{E}_{selection}(\mathcal{R}(g_0))$.  To avoid problems with numerical underflow due to small probabilities, we instead maximize the log likelihood:
\begin{equation}
\frac{1}{N}\text{ln}(\mathcal{L}(\textbf{Q})) = \frac{1}{N}\sum^N_{i=1}\text{ln} (PDF(l_i,b_i,g_i|\textbf{Q})).
\end{equation}

\subsection{Removing SDSS artifacts and globular clusters} \label{SectionCut}

In the SDSS, there are areas of the sky with either missing or unusable data, like the regions hidden behind a saturated bright star.  There are also other sections of the sky with known structures that are not included in our model like globular clusters.  We remove regions with these substructures or artifacts from our data and then integrate over the remaining usable survey volume.

The technical details of this process are as follows. We change our $PDF$ to remove the integral of the density over the volume we are cutting from the denominator of that density's respective fraction, as follows:
\begin{multline}
PDF(l,b,\mathcal{R}(g_0)|\textbf{Q}) = \\ \frac{1}{1+\sum^n_{i=1}e^{\epsilon_i}} \frac{\mathcal{E}(\mathcal{R}(g_0))\rho^{con}_{background}(l,b,\mathcal{R}(g_0)|\textbf{Q})}{\int\mathcal{E}(\mathcal{R}(g_0))\rho^{con}_{background}(l,b,\mathcal{R}(g_0)|\textbf{Q})dV - \int\mathcal{E}(\mathcal{R}(g_0))\rho^{con}_{background}(l,b,\mathcal{R}(g_0)|\textbf{Q})dV_{cut}} + \\ \sum^n_{i=1}\frac{e^{\epsilon_i}}{1+\sum^n_{j=1}e^{\epsilon_j}} \frac{\mathcal{E}(\mathcal{R}(g_0))\rho^{con}_{stream_i}(l,b,\mathcal{R}(g_0)|\textbf{Q})}{\int\mathcal{E}(\mathcal{R}(g_0))\rho^{con}_{stream_i}(l,b,\mathcal{R}(g_0)|\textbf{Q})dV - \int\mathcal{E}(\mathcal{R}(g_0))\rho^{con}_{stream_i}(l,b,\mathcal{R}(g_0)|\textbf{Q})dV_{cut}},
\end{multline}
where $dV_{cut}$ represents integrating over the volume that was cut out of the data and the rest is the same as Equation \ref{PDFEqn}.

In our data, we will either remove the stars from the globular cluster from the dataset, or we will not have data in this region due to deblending limitations in the SDSS survey.

\subsection{Covariances and degeneracies} \label{Covariances}

There are several degeneracies and covariances we must account for in our parametric model.  The parameters which are explicitly covariant are the relative weights between the density of the streams and background.  There are many other covariances that can be read from the error matrix, in Table \ref{VarianceMatrix}.  Matrix elements with a high value have high covariance in the corresponding parameters \citep{MLinAstronomy}.  When comparing two parameters, the sign of the covariance tells you the orientation of the error ellipse.  A positive covariance indicates they are positively correlated and a negative covariance indicates they are inversely correlated. An example of parameters with a high covariance is the angular position along the stripe, $\mu$, and the distance from the Sun, $R$, for the streams.

The degeneracy in stream orientation and stream selection are inherent to an optimization problem with a model of this type.  The stream orientation degeneracy comes about by the cyclic nature of angles and can be mitigated by constraining the optimization to a single hemisphere.  The stream selection degeneracy arises from the freedom to fit each stream to any location in the wedge.  Therefore, one optimization may find stream 1 to best fit Sagittarius, while another optimization on the same data may find stream 2 to best fit Sagittarius with the same parameters.  In this instance, we do not try to eliminate this degeneracy because it does not change the result in any substantive way, and attempting to force a particular stream order would require us to artificially constrain streams to mutually exclusive regions and could bias our optimization.

\section{Optimization and MilkyWay@home}

MilkyWay@home is a distributed computing platform.  Volunteers from around the world donate their unused CPU and GPU cycles to run ``workunits" in which the likelihood is evaluated for a particular set of parameters that we choose. The MilkyWay@home server, running the Berkeley Open Infrastructure for Network Computing \citep[BOINC;][]{BOINCanderson2005} sends each volunteer an executable that will calculate the likelihood, a set of model parameters to try, and one stripe of Milky Way stars, selected as in Section \ref{turnoffselection}. The users return the likelihood.  Given the newly calculated likelihood and the previously returned likelihoods, the MilkyWay@home server uses a differential evolution algorithm to determine the best parameters to send out in new workunits.

The differential evolution optimization algorithm used by MilkyWay@home is part of the Toolkit for Asynchronous Optimization (TAO) written by Travis Desell \citep{desell2007}.  This algorithm is a modified version of differential evolution for use in highly asynchronous environments such as a distributed computing platform which has heterogeneous, distributed, fault-prone hardware.  

This version of the differential evolution algorithm begins by uniformly sampling across the parameter space within the allowed search area (as determined by the constraints on each parameter). As these random sets have their likelihoods computed, it inserts them into a population.  In this context, a member of a population is a complete set of model parameters (in this case 20 parameters).  Once there are enough returned results to fill the population (for example we typically use a population size of 200, which is ten times the number of parameters being fit), newly returned likelihoods will then determine whether a member of the existing population should be replaced with the new parameter set. 

At this point, the algorithm begins trying to replace individuals within its population. To generate new parameter sets to test against the individual, the algorithm uses the following method: (1) A ``parent" is selected from the population using one of several methods.  (2) A pair of population members are randomly selected, and the differential vector between them is calculated.  Multiple pairs can be used and their differentials averaged. (3) This differential vector is scaled by a fixed differential scaling factor before adding it to the parent's parameters to identify a new position in parameter space.  Finally, (4) the individual potentially being replaced (which is in general different from the parent, and is selected by looping over the current population members) is combined in some way with the parameters of the newly determined position in parameter space.  This crossing can be done using one of several strategies, such as assigning a random chance of taking a parameter value from the new parameter set over taking a parameter value from the individual we are testing for replacement. After this crossover, the generated set of parameters obtained is sent out as a workunit.

To accommodate the asynchronous nature of MilkyWay@home, TAO blurs the lines between generations of populations in this algorithm, allowing for new parameter sets to be calculated even while waiting for results to be returned \citep{desell2007}. As results come in, the newly computed likelihood is compared with the likelihood of the population member currently residing at the ID of the individual it was attempting to replace, and the parameter set with the better likelihood is kept.  Since the lines between generations are blurred, it is possible to generate multiple test parameter sets per population member, so by the time a likelihood calculation is returned, the likelihood might be compared to that of one of the children or grandchildren of the population member from which the test parameters were originally generated.

We use a set of options that we have found consistently produce convergence within a reasonable time.  With these options, the algorithm is: (1) Parents are selected at random from the current population, which has 200 members.  (2) A single pair of population members is used to calculate the differential vector. (3) The parameter value differences are multiplied by $0.8$, and then added to the parent parameters.  (4) This point in parameter space is then combined with the parameters in the individual being tested using binary recombination, a crossover rate of 0.9, and a parent scaling factor of 1.0. 

These parameters allow the optimizers a large number of possible guesses to reduce the chance the optimizer will stall out or get stuck in a local minimum.  Typically, it takes between 2 and 3 million returned likelihood calculations to complete a single optimization of our parameters.  We often run 4 optimizations of a single data set to ensure agreement between the runs and help determine convergence.

We know a run is out of energy and converged when the change in likelihood between updates of the algorithm become small for a long period of time and the results from the independent runs on the same data agree.  When the likelihood no longer changes in the 6th decimal place for 100,000 returned results and the independent runs agree in the 6th decimal place, we assume the run is converged.  We have found that once the optimizer is making improvements on this scale, the change in the parameter values for a given improvement in likelihood is much smaller than the associated error in the parameter.

Some of the major challenges of using a volunteer distributed computing network like MilkyWay@home are: its massively asynchronous nature, its potential for faulty results, and the need to support multiple software and hardware configurations.  The project's asynchronous nature arises from the discrepancy between compute times for CPU and GPU platforms, and the episodic availability of each processor.  A CPU typically takes around 30 minutes to complete a single likelihood calculation, while a GPU can take as little as 15 seconds to complete the same calculation.  To protect our results from potential faulty likelihood calculations, we perform adaptive cross validation on our returned likelihoods.  When we receive a result, it might have to be recomputed by up to 4 more users, depending on the results of previous users. For trusted users who compute a likelihood that does not change the population, we validate a minimum of 10\% of their workunits.  The percentage validated changes based on the percentage success of their previous 10 workunits.  Results which will be integrated into the optimizer's population must be cross validated by another volunteer, who must agree on the computed likelihood for this parameter set.  If these volunteers do not agree, the work unit is sent out to additional volunteers until 2 volunteers agree on an answer. If 5 attempts to compute the likelihood do not result in two answers that agree, the parameter set is abandoned.

\section{Calculating uncertainties} \label{UncertaintySection}

We calculate parameter uncertainties using the method outlined in \cite{MLinAstronomy} for calculating uncertainties for MLEs. The uncertainty is related to the Fisher information matrix which can be calculated as the negative of the Hessian matrix of the log likelihood.  The equation for this matrix is:
\begin{equation}
H_{ij} = -\frac{d^2ln(L)}{d\theta_id\theta_j}\biggm|_{\theta=\theta_0}
\end{equation}  
where $H_{ij}$ is the Fischer information matrix. The variance matrix, $\textbf{V}$, is the inverse of the normalized Fischer information matrix:
\begin{equation}
\textbf{V} = \frac{1}{N} H_{ij}^{-1},
\end{equation}
where $N$ is the number of stars.  Then the standard deviation for each parameter is calculated by taking the square root of the diagonal elements of the variance matrix.  As mentioned in \cite{MLinAstronomy}, these are the lower bounds on the uncertainties.
When we do this calculation, we use a central finite difference around the best parameter set given by our optimizations; tests have shown that the second partial derivatives are not sensitive to the (small) stepsize.

\section{Preliminary results for SDSS stripe 19}

We fit all of the turnoff stars in stripe 19, as described in Section \ref{turnoffselection}, with a Hernquist plus double exponential disk and three streams.  We choose three streams because we expect to find, Sgr, the ``bifurcated" stream, and Virgo in our data.  However, if one of these streams is absent, the algorithm has the liberty to marginalize extra streams.  Although we plan to try fitting the data to a larger number of streams in the future, it was not attempted here due to the long processing time required to reach optimization convergence, and because the potential need to fit additional streams was not understood before completion of this work.  The parameters for the model were determined by running four full optimizations to convergence on the SDSS data, which took approximately three weeks.  The optimization with the best likelihood was selected, but the parameters and likelihoods from all four runs were similar.  

The parameters found are listed in Table \ref{ResultsTable2} along with their errors, calculated using the method described in Section \ref{UncertaintySection}.  We also provide other comparisons for our results in Table \ref{ResultsTable2} for reference.  Note that we have not used any data from neighboring stripes in creating the parameters for stripe 19; when we do the final optimizations on this data, we might find it necessary to use constraints from neighboring stripes to ensure convergence to the best possible set of parameters and to enforce the physical constraint that streams must be continuous from stripe to stripe \citep[as in][]{newby2013}, but that has not been attempted here.

A more user-friendly version of our results can be found in Table \ref{ResultsTable3}.  The parameters given in Table \ref{ResultsTable3} can be used to calculate the MSTO stellar density of the background or a stream at any point in the wedge.  The density of the smooth component, in MSTO stars per cubic kpc is given by:
\begin{equation}\label{PrelimBGDensity}
\rho_{background} = A*\rho_{smooth}(X,Y,Z)
\end{equation}
where $\rho_{smooth}(X,Y,Z)$ is from Equation \ref{background} and $A$ is provided in Table \ref{ResultsTable3}.  The density of a stream, in MSTO stars per cubic kpc is:
\begin{equation}\label{PrelimStreamDensity}
\rho_{stream} = Ae^{-r^2/2\sigma^2}
\end{equation}
where $A$ and $\sigma$ are provided in Table \ref{ResultsTable3}, and $r$ is the distance from the axis of the cylinder that describes the stream.  To calculate $r$ from the values in Table \ref{ResultsTable3}, use
\begin{equation}
r = |(\vec{r} - \vec{r_{sc}}) \times \hat{n}|
\end{equation}  
where $\vec{r}$ is the position at which you want the density and $\vec{r_{sc}}$ is the position of the stream center.  A conversion factor is required to convert from MSTO stellar density to the stellar density of other tracers based on the ratio of expected tracers to MSTO stars.
  
In Figure \ref{Stripe19Figure}, we show a wedge plot of the density of SDSS turnoff stars in stripe 19 (right panel), and the separated components (left four panels). To separate the stripe into its components, we use a probabilistic separation method.  In this separation method, we assume that each star in the stripe must belong to one of the four components we fit: the background or one of the three streams. Then, using the parts of the PDF from Equation \ref{PDFEqn} that correspond to each component, we can calculate the relative probabilities that a star belongs in each of the components.  The star is then randomly assigned to a component according to the relative probabilities.  Since this is a probabilistic separation, the specific stars we select for each component do not necessarily belong to the component in which they are placed.  The densities of the separated stars, however, follow the density of the components. For more information on this method, see \cite{cole2008} for the original derivation, and \cite{newby2013} for the derivation for multiple streams.

In this figure, you can see that the three separated streams plausibly resemble Gaussian cylinders with a slight elongation in the radial direction due to the absolute magnitude distribution, and a slight elongation due to the stream's direction of travel through the wedge; this latter elongation is not generally in the radial direction.  However, the ``bifurcated" stream is less comprehensible because it is very wide and covers the whole stripe.  The separated ``bifurcated" stream does not look like it contains any extra substructure, which would appear as substructure in the density of stars on the third panel of Figure \ref{Stripe19Figure}.  The background also looks smooth, and exhibits no apparent unidentified substructure.  As we will see, the smooth component looks very similar to the smooth component in a simulated stripe.  Figure \ref{StarFraction} compares the \cite{Xu2015} model smooth component model to our model with the parameters fit for stripe 19.  We see our model finds a lower fraction of thick disk stars in the wedge, which is expected since we designed our color cuts to eliminate most of the disk stars.

To visualize the fit the expected density distribution in this wedge, given the model and the fit parameters, we used the fit parameters to simulate each component in the SDSS wedge individually, and then combined them to get a full picture of the stripe.  The results are shown in the second row of Figure \ref{Stripe19Figure}.  The model appears to accurately fit the density of the separated components in the SDSS stripe. To get a better idea of how well it fits the SDSS data, we show a residual of the original SDSS stellar density minus the simulated density in Figure \ref{Stripe19ResidualFigure}.  We see only minor discrepancies in areas of particularly high density.

\begin{sidewaystable}\tiny	
%TO DO: Find Comparison Values for Virgo and ``bifurcated" Stream.
 \centering
 \begin{tabular}{|c |c c c c c c c c c c c c c c c c c c c c|} 
 \hline
 & \multicolumn{20}{c|}{Error Matrix for SDSS Stripe 19} \\
 \hline
 & \multicolumn{2}{c}{Background} & \multicolumn{6}{|c}{Sagittarius} & \multicolumn{6}{|c}{``bifurcated" stream} & \multicolumn{6}{|c|}{Virgo} \\
 & \multicolumn{1}{c}{$\varepsilon_{sph}$} & $q$ & \multicolumn{1}{|c}{$\varepsilon$} & $\mu$  & R & $\theta$ & $\phi$ & $\sigma$ & \multicolumn{1}{|c}{$\varepsilon$} & $\mu$ & R & $\theta$ & $\phi$ & $\sigma$ & \multicolumn{1}{|c}{$\varepsilon$} & $\mu$ & R & $\theta$ & $\phi$ & $\sigma$ \\
  & & & \multicolumn{1}{|c}{} & (deg) & (kpc) & (rad) & (rad) & (kpc) & \multicolumn{1}{|c}{} & (deg) & (kpc) & (rad) & (rad) & (kpc) & \multicolumn{1}{|c}{} & (deg) & (kpc) & (rad) & (rad) & (kpc) \\
 \hline
$\varepsilon_{sph}$ & \textbf{2.3e-7} & --- & \multicolumn{1}{|c}{---} & --- & --- & --- & --- & --- & --- & --- & --- & --- & --- & --- & --- & --- & --- & --- & --- & ---\\
$q$ & -8.2e-6 & \textbf{4.8e-4} & \multicolumn{1}{|c}{---} & --- & --- & --- & --- & --- & --- & --- & --- & --- & --- & --- & --- & --- & --- & --- & --- & ---\\
\cline{1-9}
$\varepsilon$ & 8.6e-6 & -7.2e-4 & \multicolumn{1}{|c}{\textbf{0.0077}}  & --- & --- & --- & --- & --- & \multicolumn{1}{|c}{---} & --- & --- & --- & --- & --- & --- & --- & --- & --- & --- & ---\\
$\mu$ & 4.0e-5 & -1.0e-3 & \multicolumn{1}{|c}{0.0065}  & \textbf{0.17} & --- & --- & --- & --- & \multicolumn{1}{|c}{---} & --- & --- & --- & --- & --- & --- & --- & --- & --- & --- & ---\\
R & -3.2e-6 & 6.5e-4 & \multicolumn{1}{|c}{-0.0031}  & 0.029 & \textbf{0.11} & --- & --- & --- & \multicolumn{1}{|c}{---} & --- & --- & --- & --- & --- & --- & --- & --- & --- & --- & ---\\
$\theta$ & -1.3e-5 & 6.2e-4 & \multicolumn{1}{|c}{-0.0024}  & -0.008 & 0.0042 & \textbf{0.013} & --- & --- & \multicolumn{1}{|c}{---} & --- & --- & --- & --- & --- & --- & --- & --- & --- & --- & ---\\
$\phi$ & 1.6e-5 & -7.8e-4 & \multicolumn{1}{|c}{0.0017}  & 0.018 & -0.0049 & -0.014 & \textbf{0.037} & --- & \multicolumn{1}{|c}{---} & --- & --- & --- & --- & --- & --- & --- & --- & --- & --- & ---\\
 $\sigma$ & 2.6e-6 & -3.3e-5 & \multicolumn{1}{|c}{0.0049}  & -0.0038 & 0.0032 & -0.0045 & -0.02 & \textbf{0.035} & \multicolumn{1}{|c}{---} & --- & --- & --- & --- & --- & --- & --- & --- & --- & --- & ---\\
\cline{1-1}\cline{4-15}
$\varepsilon$ & 1.2e-5 & -0.0013 & 0.0039 & 0.0044 & -0.0062 & -6.0e-4 & 7.1e-4 & 1.9e-4 & \multicolumn{1}{|c}{\textbf{0.012}} & --- & --- & --- & --- & --- & \multicolumn{1}{|c}{---} & --- & --- & --- & --- & ---\\
$\mu$ & -6.9e-4 & 0.019 & 0.003 & -0.25 & -0.29 & 0.14 & -0.13 & -0.084 & \multicolumn{1}{|c}{0.015} & \textbf{130} & --- & --- & --- & --- & \multicolumn{1}{|c}{---} & --- & --- & --- & --- & ---\\
R & 8.4e-4 & -0.033 & 0.024 & 0.29 & 0.38 & -0.14 & 0.14 & 0.079 & \multicolumn{1}{|c}{0.12} & -110 & \textbf{110} & --- & --- & --- & \multicolumn{1}{|c}{---} & --- & --- & --- & --- & ---\\
$\theta$ & -1.6e-5 & 0.001 & -0.003 & -0.0071 & -0.0015 & 0.0019 & -0.0018 & -0.0015 & \multicolumn{1}{|c}{-0.0096} & 0.64 & -0.82 & \textbf{0.017} & --- & --- & \multicolumn{1}{|c}{---} & --- & --- & --- & --- & ---\\
$\phi$ & 6.4e-6 & -2.4e-4 & 4.9e-4 & -0.001 & 0.0038 & -0.0011 & 0.0011 & 8.9e-4 & \multicolumn{1}{|c}{-0.002} & -0.72 & 0.52 & -6.4e-4 & \textbf{0.0079} & --- & \multicolumn{1}{|c}{---} & --- & --- & --- & --- & ---\\
$\sigma$ & 9.7e-5 & -0.0095 & -0.022 & 0.21 & 0.1 & 0.027 & -0.028 & -0.037 & \multicolumn{1}{|c}{0.11} & -5.9 & 11 & -0.18 & -0.038 & \textbf{6.6} & \multicolumn{1}{|c}{---} & --- & --- & --- & --- & ---\\
\cline{1-1} \cline{10-21}
$\varepsilon$ & -1.3e-6 & -6.6e-4 & 4.1e-4 & -0.0089 & 3.5e-4 & 1.2e-4 & 8.8e-4 & -0.0022 & 0.0013 & 0.16 & -0.15 & 0.0026 & 6.2e-4 & -0.012 & \multicolumn{1}{|c}{\textbf{0.0072}} & --- & --- & --- & --- & ---\\
$\mu$ & -9.0e-4 & -0.058 & 0.48 & -0.1 & -0.38 & -0.067 & 0.25 & -0.11 & 0.59 & 73 & -69 & 0.27 & -0.26 & -4.1 & \multicolumn{1}{|c}{0.81} & \textbf{280} & --- & --- & --- & ---\\
R & 3.7e-4 & -0.0068 & -0.037 & 0.0096 & -0.025 & -0.014 & -0.012 & 0.025 & -0.054 & -9.6 & 9.5 & -0.048 & 0.035 & 0.71 & \multicolumn{1}{|c}{-0.09} & -36 & \textbf{5.4} & --- & --- & ---\\
$\theta$ & 7.3e-6 & -4.4e-4 & -2.6e-4 & -0.0033 & -0.0023 & -5.2e-4 & 5.7e-4 & -3.8e-4 & 3.0e-4 & -0.039 & 0.063 & -3.3e-4 & 1.0e-4 & 0.0053 & \multicolumn{1}{|c}{7.7e-4} & -0.19 & 0.05 & \textbf{0.0013} & --- & ---\\
$\phi$ & 4.9e-6 & -4.4e-4 & 0.0019 & -9.3e-5 & -0.0035 & -6.3e-4 & 5.7e-4 & 4.6e-4 & 0.0022 & 0.19 & -0.15 & -3.1e-4 & -5.6e-4 & 0.011 & \multicolumn{1}{|c}{0.0013} & 0.71 & -0.073 & -1.1e-4 & \textbf{0.0032} & ---\\
$\sigma$ & -3.3e-5 & 0.0012 & -0.014 & -0.024 & 0.014 & 0.0021 & 0.0038 & -0.013 & -0.016 & 0.26 & -0.44 & 0.02 & 0.0028 & -0.1 & \multicolumn{1}{|c}{0.014} & 0.53 & -0.12 & 0.0012 & -0.0043 & \textbf{0.087} \\

\hline
\end{tabular}
\caption{\label{VarianceMatrix}The full variance matrix for the optimization run on the SDSS stripe 19 data.  The bold diagonal elements correspond to the reported errors ($\sigma^2$ for each parameter.  The off diagonal elements represent the covariances for each of the parameters.  We only provide the lower half of this matrix since it is symmetric.  To improve legibility, we have provided boxes which group the parameters by their corresponding streams.  Although most covariances are small, for some pairs of parameters the covariances are of the order of the error.}
\end{sidewaystable}

\begin{table}\tiny
%TO DO: Find Comparison Values for Virgo and ``bifurcated" Stream.
 \centering
 \begin{tabular}{|l c c c c c c|} 
 \hline
 \multicolumn{7}{|c|}{Preliminary Stripe 19 Results from MilkyWay@home} \\
 \hline\hline 
 Background & $\varepsilon_{sph}$ & $q$ & & & & \\ [0.5ex] 
 \hline 
 \cite{newby2013} & --- & $0.52\pm0.12$  & & & & \\
 \specialrule{.05em}{.05em}{.05em} 
 Fit &  $0.9969\pm0.0005$ & $0.57\pm0.02$ & & & & \\
 \specialrule{.05em}{.05em}{.05em} 
 Constraint &  $0.8$ to $1.0$ & $0.25$ to $1.0$ & & & & \\
 \hline \hline
 Stream & $\varepsilon$ & $\mu$ (deg) & R (kpc) & $\theta$ (rad) & $\phi$ (rad) & $\sigma$ (kpc) \\ [0.5ex] 
 \hline
 Sagittarius & $-1.94\pm0.09$ & $151.7\pm0.4$ & $21.1\pm0.3$ & $2.57\pm0.1$ & $2.78\pm0.2$ & $1.0\pm0.2$  \\
 \specialrule{.05em}{.05em}{.05em}
 Sagittarius Constraint & $-3.5$ to $0.5$ & $135$ to $172$ & $13.88$ to $32.12$ & $1.77$ to $3.03$ & $2.37$ to $3.63$ & $0.1$ to $5.0$  \\
 \specialrule{.05em}{.05em}{.05em}
 Sagittarius \citep{Belokurov2006} & --- & $152.0$  & $20\pm2$  & --- & --- & ---   \\ 
 \specialrule{.05em}{.05em}{.05em}
 Sagittarius \citep{Hernitschek2017} & --- & $168.3$  & $25.7$  & --- & --- & $5.0$   \\ 
 \specialrule{.05em}{.05em}{.05em}
 Sagittarius \citep{newby2013} & $-1.9$ & $151.0$  & $23.0$  & $2.4$ & $3.0$ & $0.9$   \\ 
 \hline 
 ``bifurcated" Stream & $-0.98\pm0.11$ & $163\pm11$ & $48\pm10$ & $1.36\pm0.13$ & $3.41\pm0.09$ & $17.6\pm2.6$  \\ 
  \specialrule{.05em}{.05em}{.05em} 
  ``bifurcated" Stream Constraint & $-2.95$ to $1.95$ & $153$ to $205$ & $22.88$ to $50.0$ & $0.00$ to $3.14$ & $2.37$ to $3.63$ & $2.0$ to $25.0$  \\
 \specialrule{.05em}{.05em}{.05em}
 ``bifurcated" Stream \citep{Belokurov2006} & --- & $175.8$  & $27\pm1$  & ---  & --- & ---  \\ 
 \specialrule{.05em}{.05em}{.05em} 
 ``bifurcated" Stream \citep{Hernitschek2017} & --- & $173$  & $17$  & ---  & --- & $10.1$  \\ 
 \specialrule{.05em}{.05em}{.05em} 
 ``bifurcated" Stream \citep{Newberg2007} & --- & $179$  & $32$  & ---  & --- & ---  \\ 
 \hline 
 Virgo & $-0.39\pm0.08$ & $206.9\pm16$ & $6\pm2.3$ & $0.50\pm0.04$ & $3.0\pm0.06$ & $6.1\pm0.3$  \\
 \specialrule{.05em}{.05em}{.05em} 
 Virgo Constraint & $-3.0$ to $3.0$ & $135$ to $230$ & $6.0$ to $26.24$ & $0.00$ to $3.14$ & $0.00$ to $3.14$ & $0.5$ to $11.0$  \\
 \hline
\end{tabular}
\caption{\label{ResultsTable2}Constraints and preliminary results from MilkyWay@home fits for SDSS stripe 19 compared to results from \cite{Belokurov2006}, \cite{Hernitschek2017}, \cite{newby2013}, and \cite{Newberg2007}.  Our results for Sagittarius are comparable with the results found in \cite{Belokurov2006}, \cite{newby2013} and in \cite{Hernitschek2017}.  The slight distance and position discrepancy in the result from \cite{Hernitschek2017} could be due to the measurement being taken $1.5^{\circ}$ outside of the plane of stripe 19. We find a much farther distance to the ``bifurcated" stream than was found in \cite{Newberg2007}, \cite{Belokurov2006} and \cite{Hernitschek2017}.   The distance discrepancy in \cite{Newberg2007} can be explained by the fact that they did not consider completeness, or the affect of larger color errors near the limit of the survey.}
\end{table}

\begin{table}
 \centering
 \begin{tabular}{|l c c c c c c c|} 
 \hline
 \multicolumn{8}{|c|}{Stripe 19} \\
 \hline \hline
 Background & Stars & A & q & $r_o$ (kpc) & $\varepsilon_{sph}$ & & \\
 \hline
 Background & $38271$ & $245099266$ & $0.57$ & $12.0$ & $0.9969$ & & \\
 \hline \hline
 Stream & Stars & l (deg) & b (deg) & r (kpc) & A & $\sigma$ (kpc) & $\hat{n}$ \\ [0.5ex] 
 \hline
 Sagittarius & $5500$ & $215.5$ & $49.2$ & $21.1$ & $1143.4$ & $1.00$ & $(-0.51,  0.19, -0.84)$ \\ 
 \hline 
 ``bifurcated" Stream & $14363$ & $218.6$ & $60.8$ & $48.2$ & $58.4$ & $17.6$ & $(-0.94, -0.26,  0.21)$ \\ 
 \hline 
 ``Virgo" & $25911$ & $12.4$ & $74.4$ & $6.05$ & $360.9$ & $6.12$ & $(-0.48,  0.06,  0.88)$ \\ 
 \hline
\end{tabular}
\caption{\label{ResultsTable3}Easy-to-use description of the density of MSTO stars in the background and streams, the preliminary results in Table \ref{ResultsTable2}. This table includes the number of stars in the wedge, converted from the weights; the stream center's Galactic l, b, and r, converted from $\mu$, $\nu$ and $r$; the stream's Gaussian normalization factor A and width $\sigma$; and the stream's unit direction in Galactic $(X,Y,Z )$.  With the numbers provided in this table it is possible to find the background and stream MSTO star densities at any point the wedge using Equations \ref{PrelimBGDensity} and \ref{PrelimStreamDensity}.}
\end{table}

\section{Validation of the results using test data}

\subsection{Generating simulated turnoff stars in SDSS stripe 19} \label{GenTestData}

To test that both the model and fitting algorithm work as intended, the algorithm was tested on a set of data drawn from a density model with known parameters.  We wrote a completely separate program to generate the test data, so that the entire modeling and fitting procedure would be tested.  Generating the test data is a multi-step process in which we simulate our background stellar distribution and streams with the correct number of stars, determine the star's absolute and apparent magnitudes using our magnitude distribution, and then apply the effects of observational bias.

The first step to generating our test data in selecting the set of parameters we wish to simulate.  To do this, we use the values of the parameters from our fit to SDSS stripe 19 from Table \ref{ResultsTable2}.  This is the best set of parameters to check the accuracy of the derived stripe 19 density model.

After we choose the parameters, we input them to our wedge simulator, which begins by calculating the number of stars in each component of our data.  The program does this using Equation \ref{weights} with the specified weights to get the fraction of stars in each component.  The fraction of stars is then multiplied by the number of MSTO stars we want to simulate (in this case 84,046, which is the number of turnoff stars observed in stripe 19).  Each component is then simulated with the required number of stars.

For each component of the density, the test data generator creates one star at a time until the number required by the weights is reached.  The smooth component is generated using rejection sampling.  We uniformly sample Equation \ref{background} over the volume of the wedge, and we compare the density from Equation \ref{background} to the maximum density possible in the wedge from Equation \ref{background}.  Then, using a random number, we determine whether we will reject or keep the star, based on the fraction of the calculated density to the maximum density.  

Stream stars are generated using active generation to create a stellar position in three dimensions. One number drawn from a uniform distribution determines the position along the cylinder's z-axis, and two numbers drawn from a normal distribution determine the position of the star in the plane perpendicular to the cylinder's z-axis.  The length of the cylindrical distribution that is generated is significantly longer than the portion of the cylinder axis that is within the wedge of data in stripe 19.  The cylinder's length is long enough that all of the volume in the wedge that is within three $\sigma$ of the cylinder's z-axis will be populated.  The stream star is then rotated and translated into Galactic X, Y, Z based on the stream's orientation and center point. 

As the program generates each star, it applies the observational biases that are corrected for in our maximum likelihood model.  First, it determines the star's absolute magnitude based on the absolute magnitude distribution in Equation \ref{Modfit} and then uses that to determine its apparent magnitude.  At this point it checks to make sure the star lies within the wedge.  Finally, it samples the combined magnitude limit from Equation \ref{sig} and the color selection efficiency from Equation \ref{starleak} to determine if the star would have been observed.

\subsection{Results from fitting test data}

We generated two different sets of test data: one with a Hernquist background that exactly matches the model to which the data is fit, and one with a broken power law background. The second simulation tests the sensitivity of our measurements of the stream parameters to imperfect knowledge of the smooth component of the Milky Way spheroid.  For the broken power law background, we used the parameters fit in \cite{shaila2012}: an inner halo exponent of $-2.78$, outer halo exponent of $-5.0$ and break radius of $45$ kpc.

MilkyWay@home was used to find the optimum values of the parameters, fit separately for the two sets of generated test data.  We ran 4 optimizations for each of the simulated data wedges, each with minimal constraints.  Each set of 4 optimizations independently converged to the same results; the results are listed in Table \ref{ResultsTable}.  

Next, we determined the uncertainty in each parameter using the method outlined in Section \ref{UncertaintySection}.  Since our errors are given as one standard deviation, we can expect 68\% (13-14 parameters) of the results to be within uncertainty, 95\% (19 parameters) within two times the uncertainty and the rest to be within three times the uncertainty most of the time.  Our results in Table \ref{ResultsTable} show that for the Hernquist plus disk background simulation, we had 10 parameters within 1 times the uncertainties, 16 within 2 times the uncertainties, and 18 were within 3 times the uncertainties.  While it is unusual that two of the parameters are more than 3 times the uncertainty from their simulated value, we do not believe this is enough to cause alarm.

In our results for the simulation with the broken power law model, we find that 5 parameters are within 1 times the uncertainties, 8 are within 2 times the uncertainties and 9 are within 3 times the uncertainties.  The higher frequency of best fit parameters with large offsets from the simulated values shows that if our model is not a good match to the actual stellar density model in the halo, our errors, as calculated using the method in Section \ref{UncertaintySection}, could be underestimated by a factor of three or more.  The fraction of stars in each component is particularly poorly measured compared to the uncertainties; the optimizations for $\varepsilon$ differ from the simulated values by 6 to 11$\sigma$.  These errors look worse than they really are because they are measured with respect to a different smooth component.  We include a column in Table \ref{ResultsTable} that shows the number of stars in the stripe associated with each component.  The density of stars along a stream could be off by as much as 60 percent.

Nevertheless, the properties of the detected streams and the general shape of the stellar spheroid (as measured by $q$) are generally similar to the simulation.  All of the halo substructure is immediately recognizable.  The flattening of the spheroid is surprisingly correct.  The widths of the streams are approximately correct, the stream centers and distances are pretty close, and even the angles at which the steams pass through the wedge are easily matched with the correct stream.

In addition to checking the accuracy of the uncertainties, we checked the accuracy of the optimizer in finding the highest point in the likelihood surface.  To do this, we ran one dimensional parameter sweeps for each of the 20 parameters in our model.  The sweeps seen in Figure \ref{Full1DSweepSim} show the likelihood surface for the test data with the Hernquist background.  This figure shows that each model parameter peaks either at, or close to, the expected parameter value.  Any small difference between the peak in the likelihood surface and the simulated values can be accounted for by the uncertainty introduced by the finite number of stars available in our data.  For the broken power law parameter sweeps in Figure \ref{Full1DSweepSimBPL}, there are several parameters which have extremely flat likelihood surfaces.  In these cases, the optimizer still seems to find an answer close to the peak of the likelihood.

In order to visually compare the results of our simulation to the actual SDSS stripe 19 data (Figure \ref{Stripe19Figure}), we present the simulated data with a Hernquist background in Figure \ref{Sim19Figure}, and the simulated data with the broken power law background in Figure \ref{Sim19BPLFigure}.  The simulated data figures, Figure \ref{Sim19Figure} and Figure \ref{Sim19BPLFigure}, show three different views of the data.  The first row shows the components generated by the test data simulator, and the combination of those components.  The second row shows a probabilistic separation of the components by our model, given the simulation parameters.  Finally, the last row shows a re-simulation of the data using the fit model parameters.  By re-simulating the stripe, we can visualize the model that the optimizer found as the best fit to the data.

In each of these plots the separated streams have the appearance of streams, and the smooth component appropriately contains no substructure.  Especially interesting is the ability for MilkyWay@home to determine, on its own, a set of parameters that resulted in streams that look, by eye, the same as those with the ``correct" parameters for the wedge.  More evidence of this can be seen in Figures \ref{Sim19ResidualFigure} and \ref{Sim19ResidualBPLFigure}, which show residual densities of the simulated data and re-simulated fit data from Figures \ref{Sim19Figure} and \ref{Sim19BPLFigure}.  The broken power law fits show that we will not easily know whether or not the real data is a poor fit to a Hernquist background.  However, we will still be able to fit most stream parameters (8 of the 18 stream parameters were within 2 sigma of the simulated values), even though a few might not be correct within the calculated errors.

In summary, simulations show that in the case that our background and stream models are close to the correct answer, the resultant parameter fits can be trusted.  If our background or stream models are far from the correct distribution of stars the uncertainties are underestimated, but the stream properties are generally reasonable.

\setlength{\arrayrulewidth}{.5mm}
\begin{table}\tiny
\centering
 \begin{tabular}{|l c c c c c c c|} 
 \hline
 \multicolumn{8}{|c|}{Simulated and Fit Values from MilkyWay@home} \\
 \hline\hline 
 Background & $N_{stars}$ & $\varepsilon_{sph}$ & $q$ & & & & \\ [0.5ex] 
 \hline 
 \textit{Simulated Value} & $\textit{38,270}$ & $\textit{0.997}$ & $\textit{0.565}$ & & & & \\
 \specialrule{.05em}{.05em}{.05em} 
 Fit & $37,752\pm1,586$ & $0.997\pm0.0004$ & $0.54\pm.02$  & & & & \\
 \specialrule{.05em}{.05em}{.05em} 
 BPL & $29,431\pm853$ & $1.0\pm0.0003$ & $0.58\pm.02$  & & & & \\
 \specialrule{.05em}{.05em}{.05em} 
 Constraint & --- & $0.8$ to $1.0$ & $0.25$ to $1.0$ & & & & \\
 \hline \hline
 Stream & $N_{stars}$ & $\varepsilon$ & $\mu$ (deg) & R (kpc) & $\theta$ (rad) & $\phi$ (rad) & $\sigma$ (kpc) \\ [0.5ex] 
 \hline
 \textit{Sagittarius (Sim Value)} & $\textit{5,500}$ & $\textit{-1.94}$ & $\textit{151.7}$ & $\textit{21.1}$ & $\textit{2.57}$ & $\textit{2.78}$ & $\textit{1.0}$  \\ 
 \specialrule{.05em}{.05em}{.05em} 
 Sagittarius (Fit) & $5,877\pm759$ & $-1.86\pm0.08$ & $152.0\pm0.4$ & $20.7\pm0.3$ & $2.43\pm0.10$ & $2.80\pm0.14$ & $1.2\pm0.2$ \\ 
 \specialrule{.05em}{.05em}{.05em} 
 Sagittarius (BPL) & $8,776\pm711$ & $-1.21\pm0.07$ & $151.7\pm0.4$ & $21.5\pm0.3$ & $2.47\pm0.10$ & $2.43\pm0.22$ & $1.9\pm0.4$ \\ 
 \specialrule{.05em}{.05em}{.05em}
 Sagittarius Constraint & --- & $-3.5$ to $0.5$ & $135$ to $172$ & $13.88$ to $32.12$ & $1.77$ to $3.03$ & $2.37$ to $3.63$ & $0.1$ to $5.0$  \\
 \hline
 \textit{``bifurcated" Stream (Sim Value)} & $\textit{14,363}$ & $\textit{-0.98}$ & $\textit{163.2}$ & $\textit{48.2}$ & $\textit{1.36}$ & $\textit{3.41}$ & $\textit{17.6}$  \\ 
 \specialrule{.05em}{.05em}{.05em} 
 ``bifurcated" Stream  (Fit) & $18,193\pm1,641$ & $-0.73\pm0.1$ & $180.0\pm5$ & $36\pm3$ & $1.31\pm0.12$ & $3.27\pm0.10$ & $16.1\pm2.1$  \\ 
 \specialrule{.05em}{.05em}{.05em} 
 ``bifurcated" Stream  (BPL) & $18,395\pm1,154$ & $-0.47\pm0.07$ & $204.6\pm13$ & $32\pm0.8$ & $1.31\pm0.06$ & $3.63\pm0.22$ & $15.6\pm1.8$  \\ 
  \specialrule{.05em}{.05em}{.05em} 
  ``bifurcated" Stream Constraint & --- & $-2.95$ to $1.95$ & $153$ to $205$ & $22.88$ to $50.0$ & $0.00$ to $3.14$ & $2.37$ to $3.63$ & $2.0$ to $25.0$  \\
 \hline 
 \textit{Virgo (Sim Value)} & $\textit{25,911}$ & $\textit{-0.39}$ & $\textit{206.9}$ & $\textit{6.05}$ & $\textit{0.50}$ & $\textit{3.01}$ & $\textit{6.12}$  \\ 
 \specialrule{.05em}{.05em}{.05em} 
 Virgo (Fit) & $22,221\pm1,641$ & $-0.53\pm0.08$ & $214.9\pm9$ & $6.1\pm0.7$ & $0.51\pm0.03$ & $3.13\pm0.04$ & $5.5\pm0.3$  \\
 \specialrule{.05em}{.05em}{.05em} 
 Virgo (BPL) & $27,441\pm1,295$ & $-0.07\pm0.06$ & $194.1\pm4$ & $6\pm1$ & $0.42\pm0.02$ & $3.11\pm0.03$ & $4.9\pm0.2$  \\
 \specialrule{.05em}{.05em}{.05em} 
 Virgo Constraint & --- & $-3.0$ to $3.0$ & $135$ to $230$ & $6.0$ to $26.24$ & $0.00$ to $3.14$ & $0.00$ to $3.14$ & $0.5$ to $11.0$  \\
 \hline
\end{tabular}
\caption{\label{ResultsTable}Results and constraints of the MilkyWay@home fits for simulated data wedges with Hernquist (Fit) and Broken Power Law (BPL) backgrounds.  Here we can see that most of the simulated parameters lie within our one sigma errors from our fits, when the simulated density has the same form as the model to which it is fit (Fit).  The parameter values for the BPL fits are generally similar to the simulated values, except some of the uncertainties seem to be underestimated using our uncertainty estimation method.}
\end{table}

\section{Discussion}

Given that our tests support the validity of our algorithm in determining the properties of large halo substructures, we now discuss the results obtained from SDSS stripe 19.

Table \ref{ResultsTable2} shows the similarity between our measurement of the Sgr dwarf tidal stream leading tidal tail and those of \cite{Belokurov2006} and \cite{newby2013}.  The larger difference between our position and that of \cite{Hernitschek2017} can be explained by the fact that their position is not exactly aligned with stripe 19; the \cite{Hernitschek2017} results are about a degree and a half away from ours in $\nu$.  \cite{newby2013} found that the stream position can vary up to $11$ degrees in $\mu$ and up to $3$ kpc in distance between $2.5$ degree stripes. The discrepancies between the Sgr widths in \cite{Hernitschek2017} and our results can be explained by their different measurement methods.  The \cite{Hernitschek2017} widths are based on line-of-sight depth, which is prone to uncertainty from stellar distance uncertainty and stream orientation, making this likely an overestimate of width.  In addition, our Sgr measurements are consistent with positions and distances reported in N-body simulations and fits to data from sources like \cite{Law2005}, \cite{Belokurov2006}, \cite{Newberg2007}, and \cite{Law2010} which all report a Sgr leading arm distance between 20 and 30 kpc near the SDSS stripe 19 data wedge.

The ``bifurcated" stream results advance our knowledge of the Sgr stream.  Our results are inconsistent with those reported in \cite{Hernitschek2017}.  They report a distance of $17$ kpc, which is quite different from our result of $48$ kpc. Comparing our results with the \cite{Newberg2007} results in Table \ref{ResultsTable2}, we find our stream center results are close enough that the differences, approximately 1.5 times our one sigma uncertainties, could be purely statistical. However, we believe the results in this paper are more accurate because \cite{Newberg2007} did not account for the effects of completeness or the effects of stars leaking into/out of the color selection box when calculating distance. By neglecting these biases in the data, Newberg et al. underestimated the distance to the ``bifurcated" stream.  Our correlation (as discussed in Section \ref{Covariances} and given in Table \ref{VarianceMatrix}) suggests that this would result in a measurement of $\mu$ that is too large, which explains why our measured $\mu$ is smaller.  Our ``bifurcated" stream center is more consistent with those of the Sgr trailing tail in the north Galactic cap, as measured in \cite{Belokurov2006} (as branch C) and \cite{Belokurov2014}, and as simulated by \cite{Law2005} and \cite{Law2010}. From these sources, the Sgr trailing tidal tail is expected to be between $40$ and $65$ kpc, which is consistent with our result.  This begs the question of where the ``bifurcated" stream is in this wedge.  It is possible that the ``bifurcated" stream in the \cite{Belokurov2006} Field of Streams image is the Sgr trailing tail.  Previous measurements of the ``bifurcated" stream at the distance of the leading tidal tail could come from the leading tidal tail itself.  Alternatively, there could be a ``bifurcated" stream in the stripe 19 data wedge that was not fit by any of the three streams.  More work is required to solve this ambiguity.

If we are characterizing the trailing tail of Sgr, then our width measurements could have implications for the shape and properties of the Milky Way dark matter halo \citep{Ibata2001, SiegalGaskinsValluri2008, Ngan2016, Sandford2017}. \cite{Ibata2001} find that a wide tail for a stellar stream is indicative of a flattened or oblate halo.  \cite{SiegalGaskinsValluri2008}, \cite{Ngan2016}, and \cite{Sandford2017}, present methods to determine the number of dark matter subhalos and constraints on their mass by looking at the width and gaps present in the tidal streams.  In general, more gaps and wider streams seem to correlate with larger, and higher numbers of subhalos in the dark matter halo.  Since we measure a width of $17.6$ kpc for the trailing tail of Sgr it could suggest an oblate or a lumpy dark matter halo.  Our measurements suggest a stream that is a $47^{\circ}$ FWHM across, at an Sgr lamda of $128^{\circ}$. Since lamda increases in the direction of the leading tidal tail, starting at the dwarf galaxy, this is $232^{\circ}$ along the trailing tidal tail.  For more information on the Sagittarius coordinate system, see \cite{Majewski2003}.

The distance to the assumed Virgo Overdensity in Table \ref{ResultsTable2} is consistent with, but on the low end of, the distance found in \cite{Juric2008} of $6$-$20$ kpc.  \cite{Juric2008} also suggest that Virgo is over $1000$ square degrees on the sky, which is consistent with our close and wide model parameters of a $6$ kpc distance and $6$ kpc width.  A sigma of $6$ kpc corresponds to a $14$ kpc FWHM, which at $6$ kpc away is $67^{\circ}$ across.  A circular region of the sky with this width is 3300 sq. deg.  Note, however that our width is from a cross section near the edge of the overdensity, which makes this calculation quite uncertain.  Also note that this observed substructure is close enough and wide enough that we might expect to see some stars associated with this stream in the solar neighborhood.  \cite{Duffau2006} also suggest the stream is wide, taking up over 100 square degrees on the sky; they find a much more distant substructure.  \cite{CarlinVirgo} fit an orbit to the Virgo stellar stream and present an N-body simulation of the stream which shows two wraps of Virgo in the region we fit.  The distances to the trailing wrap of the stream's orbit are consistent with our fits to the Virgo Overdensity, while the leading wrap is closer to $15$ kpc.  The Virgo Overdensity ``stream center" we fit is several degrees away from the plane of the orbit and has a higher density than the N-bodies suggest.  Our fits are close enough that we can not rule our the possibility that our structure is associated with this fit of the Virgo stellar stream, but we also can not conclusively confirm that what we are seeing is the Virgo Overdensity tidal stream as outlined by \cite{CarlinVirgo}.  Future analyses of the density substructure of the north Galactic cap will map the MSTO density in the Virgo region, allowing us to address the question of whether this large overdensity consists of multiple tidal streams.

\section{Conclusions}

\begin{figure}
\centering
\includegraphics[width=1.0\textwidth]{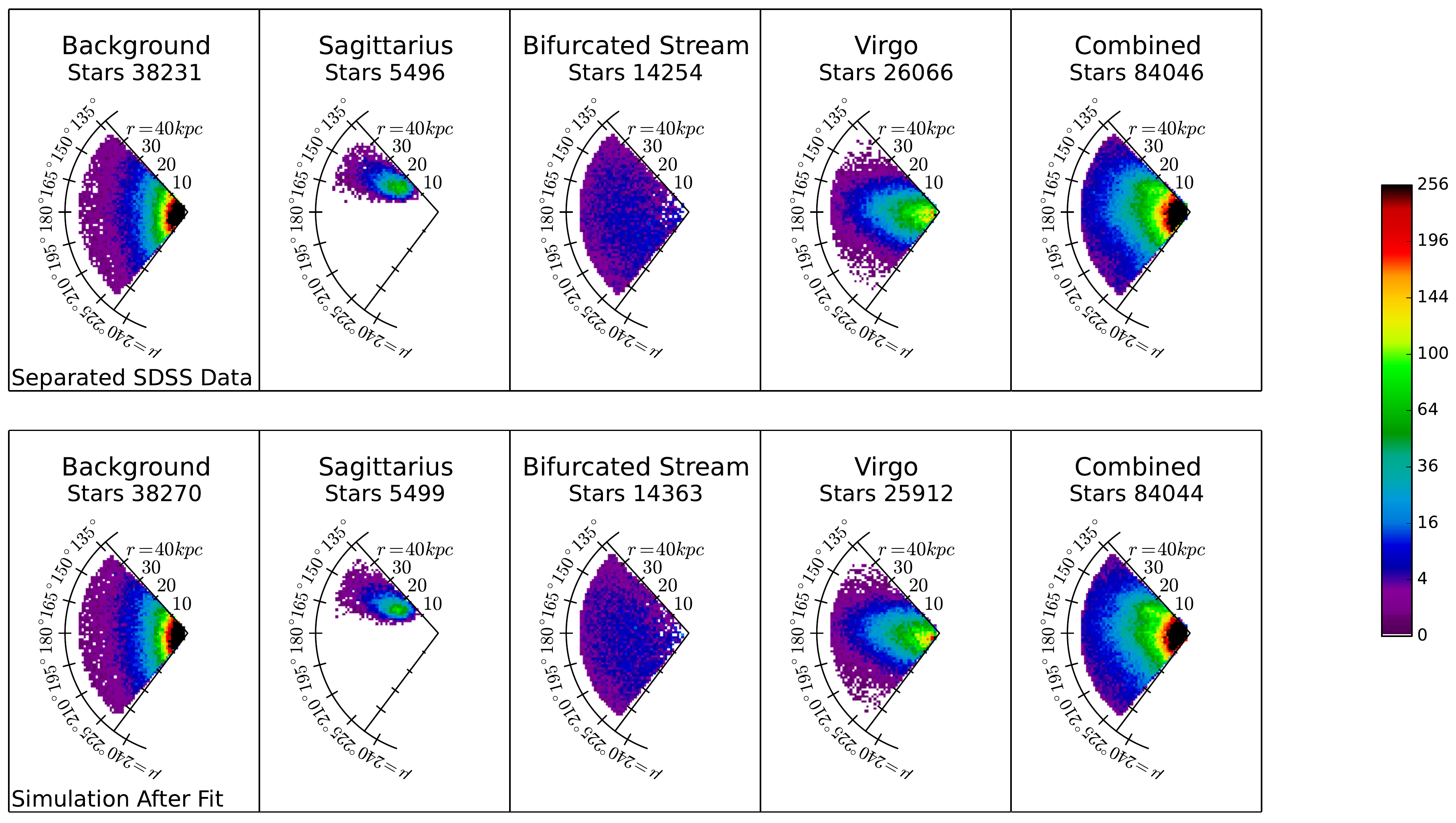}
\caption{MSTO star stellar density in stars per cubic kpc in a 1 kpc by 1 kpc by 2.5 degree pixel at each point in a flattened, face on view of SDSS stripe 19 after they were fit by MilkyWay@home.  The first four panels in the top row show a probabilistic separation of the SDSS stars into each different substructure fit, and the last panel includes all of the stars in the original stripe of data before separation. In each plot, we show a smoothly separated background, and three separate streams with seemingly no substructure left over in the wedge.  The lower row shows the density distribution of stars in a simulated stripe, that was created with the same parameters that were fit to the actual data.  Note the similarity between the simulation with three streams and a smooth background to the actual SDSS data.}
\label{Stripe19Figure}
\end{figure}

\begin{figure}
\centering
\includegraphics[width=0.33\textwidth]{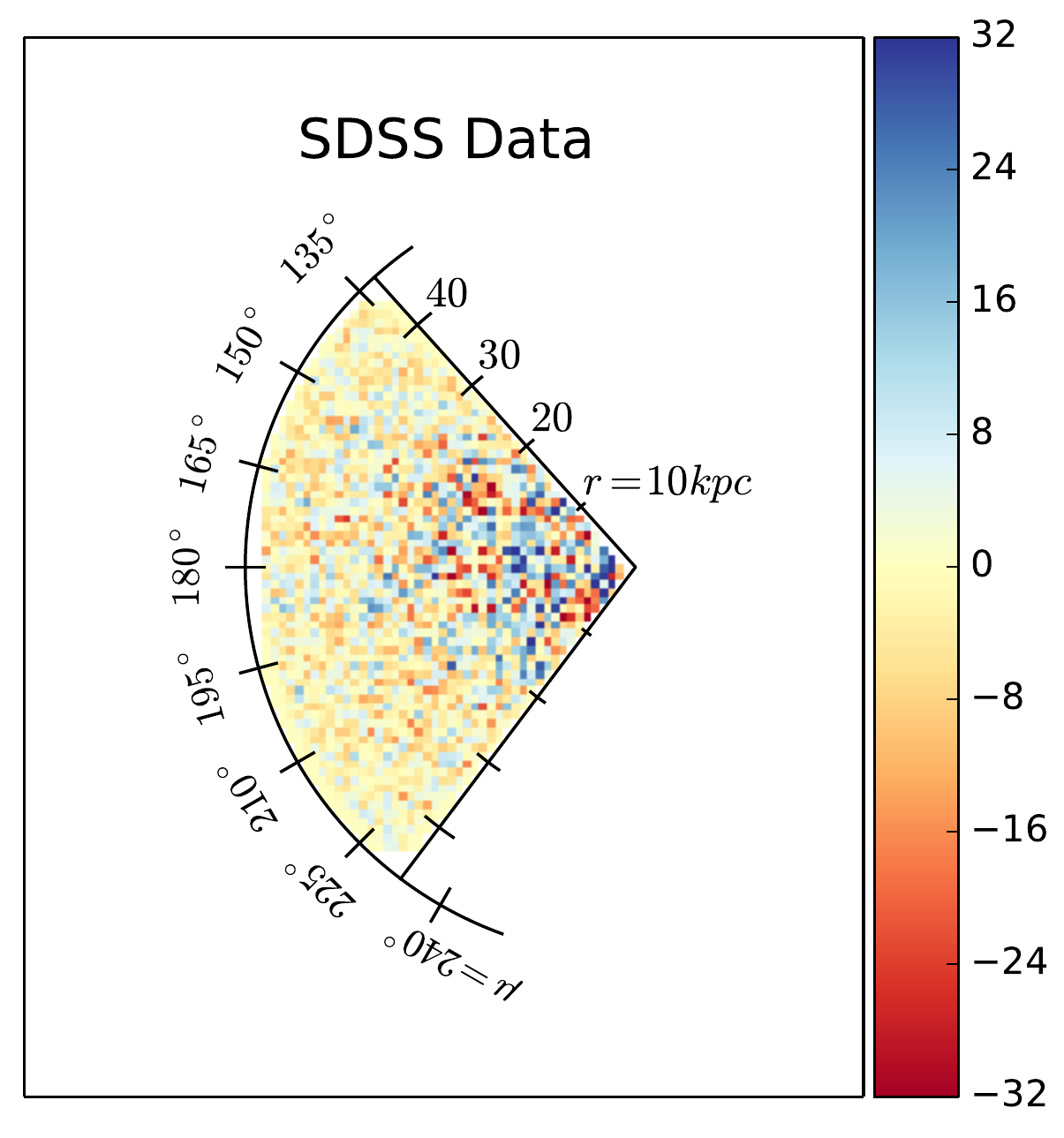}
\caption{The residual between the stellar density of SDSS stars in stripe 19 and a simulation created with the fit parameters from MilkyWay@home. The color represents the residual density of MSTO stars per cubic kpc in a 1 kpc by 1 kpc by $2.5^{\circ}$ pixel in the flattened (all stars are added together in the $\nu$ direction, so the volume of the stripe increases with radius from the Sun), face-on stripe.  This residual is found by subtracting the ``Simulation After Fit" data from the SDSS data from Figure \ref{Stripe19Figure}.  The residual shows that the model fits the data well, though there remains some structure in the highest density portion of the residual.}
\label{Stripe19ResidualFigure}
\end{figure}

\begin{figure}
\centering
\includegraphics[width=1.0\textwidth]{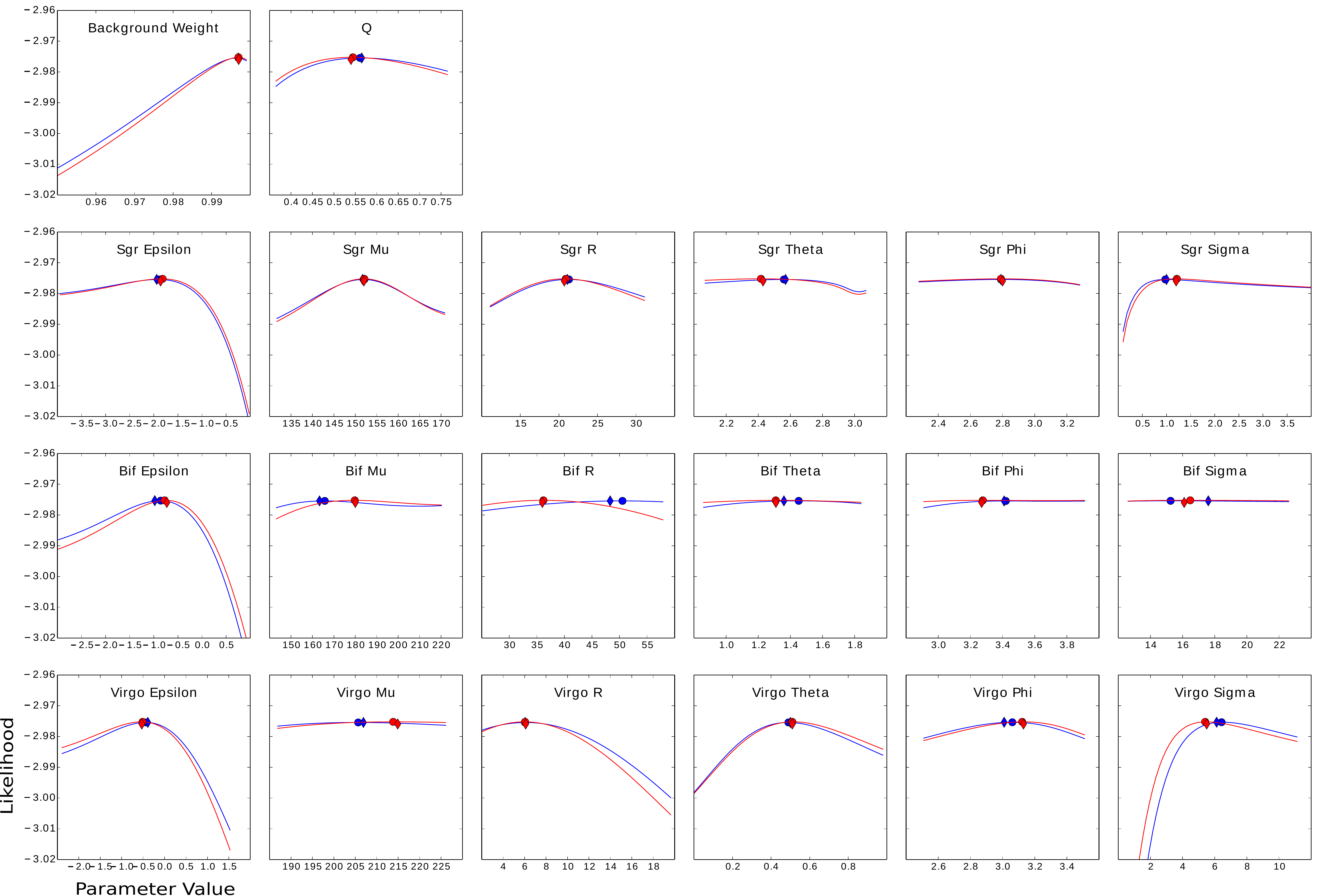}
\caption{Parameter sweeps in each of the 20 parameters for the simulated SDSS stripe 19 with the Hernquist plus thick disk background.  A parameter sweep gives the $log(\mathcal{L})$ as a function of one of the variables, holding the other variables constant.  The blue line represents a parameter sweep around the simulated parameters from Table \ref{ResultsTable}, and the red line represents the parameter sweep around the best results returned from MilkyWay@home, listed as ``Fit" in Table \ref{ResultsTable}.  The simulated value for the parameter is shown by the blue diamond and the best likelihood in the parameter sweep around the simulated parameters is shown by the blue circle.  Similarly, the red diamond indicates the value of the parameter returned by MilkyWay@home and the red circle represents the best likelihood found in the parameter sweep around the result returned from MilkyWay@home.  In the first row, we show parameter sweeps for the smooth background parameters.  The next three rows show the parameters for, the Sagittarius stream (Sgr), the ``bifurcated" stream (Bif), and Virgo, respectively.  The best results will be produced if the parameter sweeps show a narrowly peaked likelihood surface.  Most of the panels show well-behaved slices through the likelihood surface, which is good.  All of the sweeps show that both the returned result from MilkyWay@home and the parameter sweep have the best likelihood in the same place.  This means our optimizer successfully converged to a maximum in our likelihood surface.}
\label{Full1DSweepSim}
\end{figure}

\begin{figure}
\centering
\includegraphics[width=1.0\textwidth]{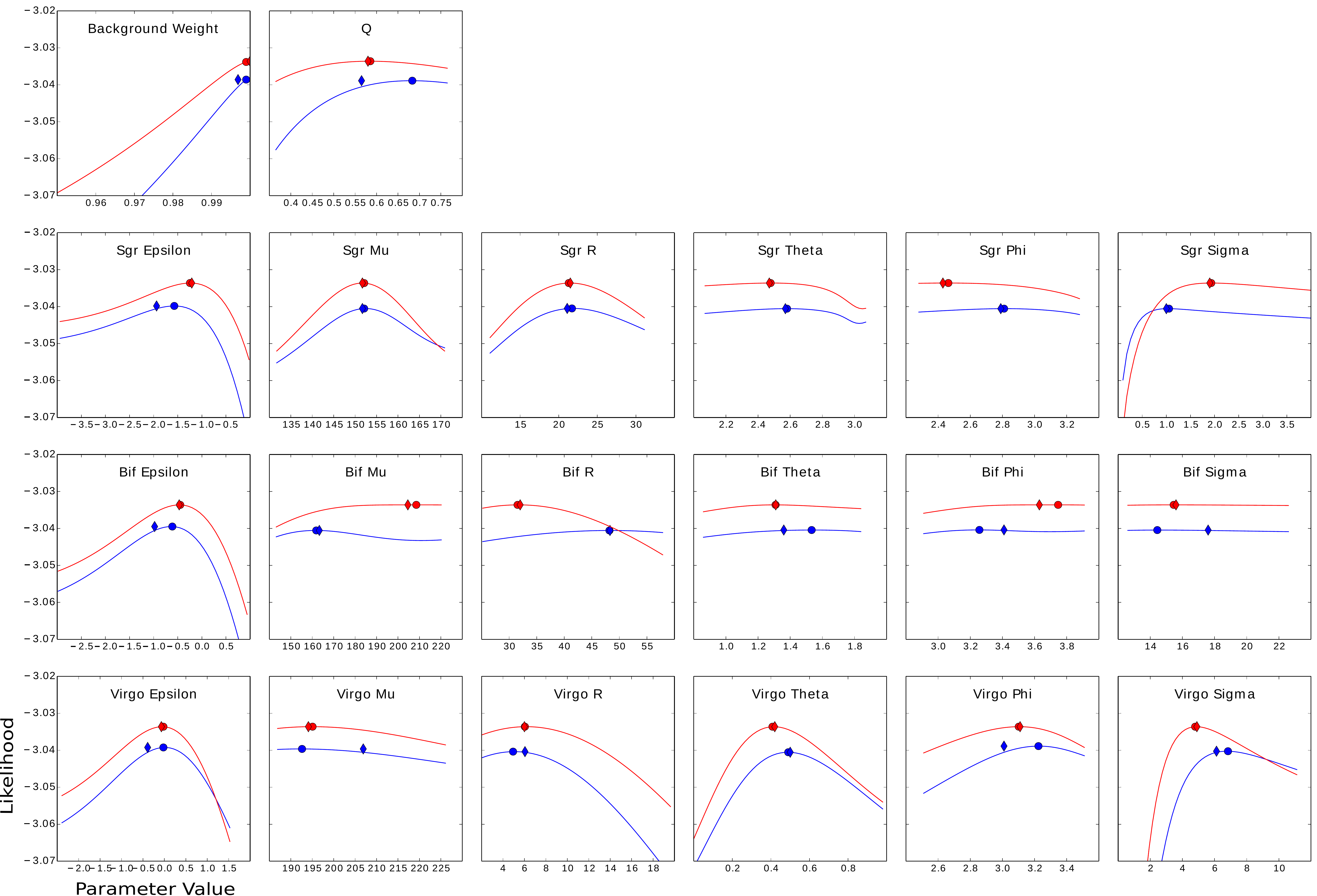}
\caption{Parameter sweeps in each of the 20 parameters for the simulated SDSS stripe 19 with the broken power law background.  The blue line represents a parameter sweep around the simulated parameters from Table \ref{ResultsTable}, and the red line represents the parameter sweep around the best results returned from MilkyWay@home, listed as ``BPL" in Table \ref{ResultsTable}.  The simulated value for the parameter is shown by the blue diamond and the best likelihood in the parameter sweep around the simulated parameters is shown by the blue circle.  Similarly, the red diamond indicates the value of the parameter returned by MilkyWay@home and the red circle represents the best likelihood found in the parameter sweep around the result returned from MilkyWay@home.  In the first row, we show parameter sweeps for the smooth background parameters.  The next three rows show the parameters for: the Sagittarius stream (Sgr), the ``bifurcated" stream (Bif), and Virgo, respectively.  Again, most of the panels show well-behaved slices through the likelihood surface.  In this set of parameter sweeps, we also note the likelihood surface around several parameters in the ``Bif" row are very flat and the optimizer was still capable of finding the peaks.}
\label{Full1DSweepSimBPL}
\end{figure}

\begin{figure}
\centering
\includegraphics[width=0.99\textwidth]{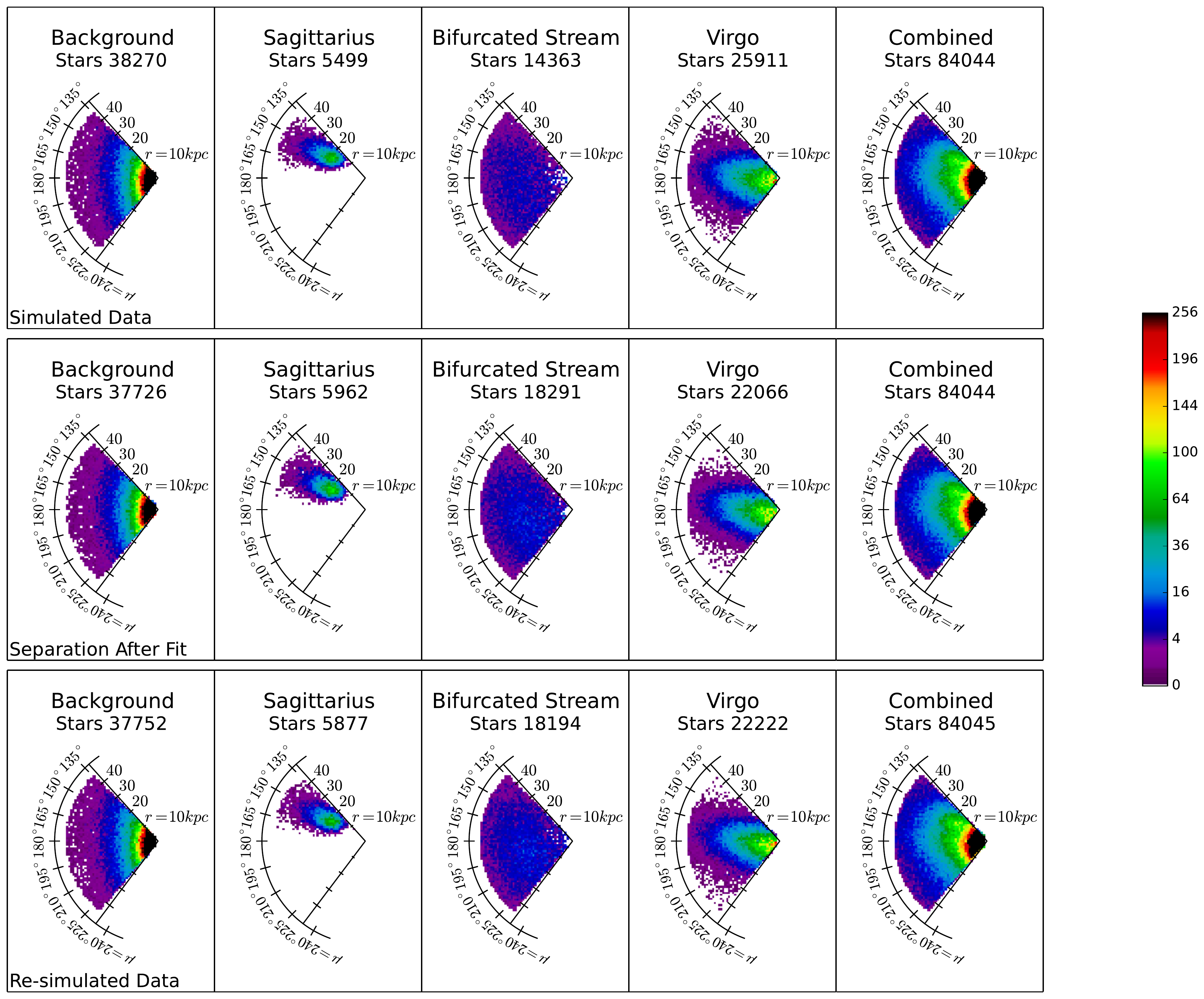}
\caption{Stellar density from our simulated stripe 19 plotted in three different ways.  In this figure, the color represents the density of MSTO stars per cubic kpc in a 1 kpc by 1 kpc by 2.5 degree pixel in the flattened, (all stars are added together in the $\nu$ direction, so the volume of the stripe increases with radius from the Sun) face-on stripe.  Values of $\mu$ and $r$ in each stripe are indicated.  In all three rows, the last panel shows the sum of the components in the first four panels.  We show the smooth component of the spheroid (background), the Sagittarius stream, the ``bifurcated" stream, and the Virgo Overdensity.  In the first row, the first four panels are the model components that are added together to make the simulated stripes.  In the second row, the first four panels are a probabilistic separation of the stars into each of the model components using the ``correct" parameters used to simulate the wedge.  Finally, in the last row, we re-simulate the wedge using the parameters recovered from the MilkyWay@home results and then separate the re-simulated wedge into its model components.  Using this method, we can visualize the model that the optimizer thinks is the best fit to the data. All three methods yield similar results for each stream, which is as expected if the separation algorithm is successful.}
\label{Sim19Figure}
\end{figure}

\begin{figure}
\centering
\includegraphics[width=0.99\textwidth]{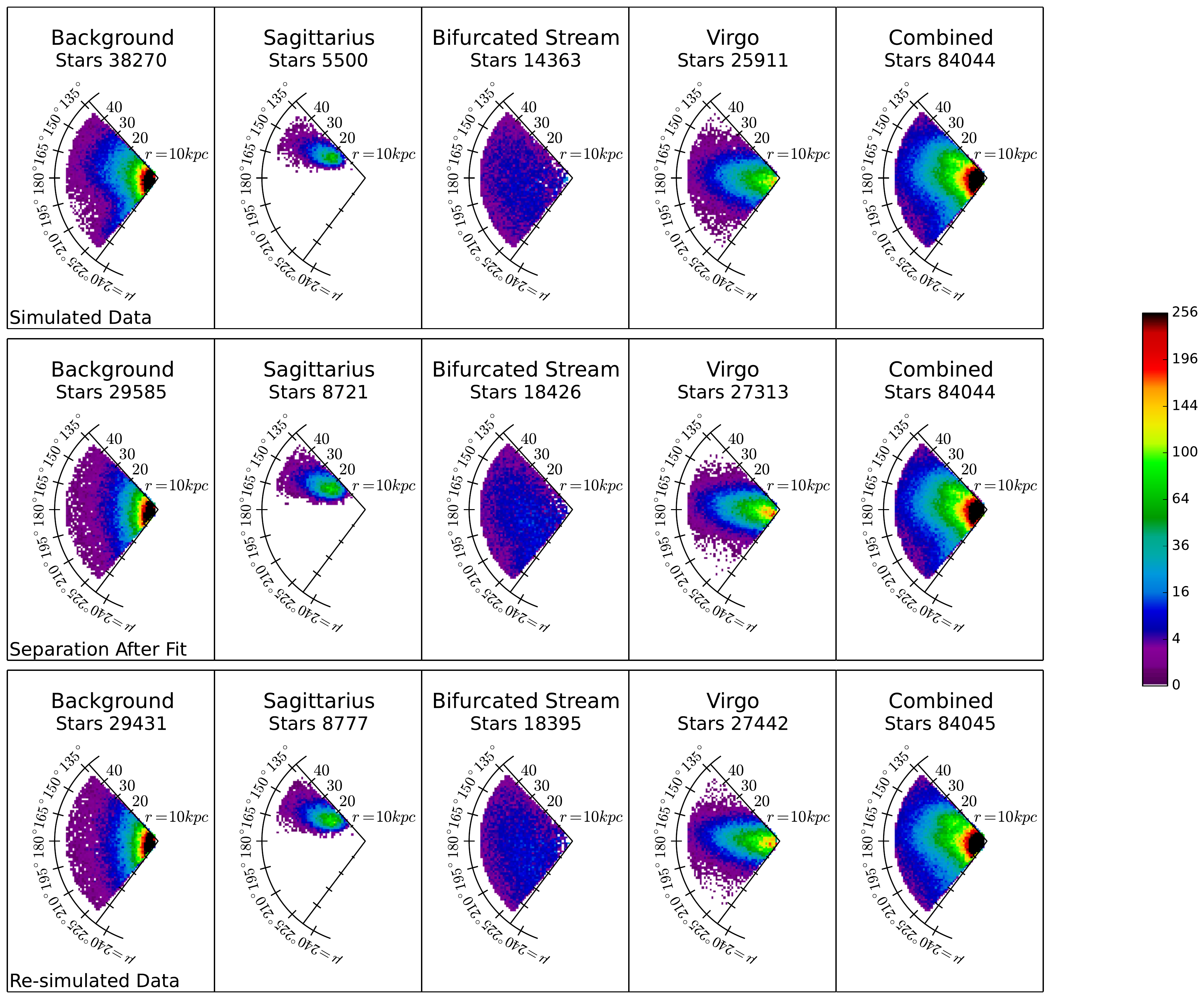}
\caption{Stellar density from our simulated stripe 19 with a broken power law background.  This figure is similar to Figure \ref{Sim19Figure}, except using a different simulated data set.  In the first row, the first four panels are the model components, including the broken power law background, that make up the simulated stripes.  In the second row, the first four panels are a probabilistic separation of the stars into each of the model components using the ``correct" parameters used to simulate the wedge.  This separation assumes a Hernquist background with parameters like those from our other simulated wedge.  Finally, in the last row, we re-simulate the stripe using the parameters recovered by MilkyWay@home with a Hernquist background.  The first four panels are the four components and the last panel is a combination of all of the stars.  By re-simulating the stripe using the recovered model parameters, we can visualize the model that the optimization found as the best fit to the data.  Note that MilkyWay@home is not able to faithfully reproduce the simulated background if the model that is being fit is not the same as actual stellar halo density distribution. This is evident from the fact that the shape of the smooth component in the upper left panel shows significant extra structure at $\mu = 225^{\circ}$ and much less at $\mu = 210^{\circ}$.  Since the Hernquist profile cannot reproduce this shape, the separation in the lower left splits the difference between these two densities; the streams are shaped slightly differently to absorb or contribute stars as necessary.  Note that overall, the stream densities are remarkably robust. }
\label{Sim19BPLFigure}
\end{figure}

\begin{figure}
\centering
\includegraphics[width=0.33\textwidth]{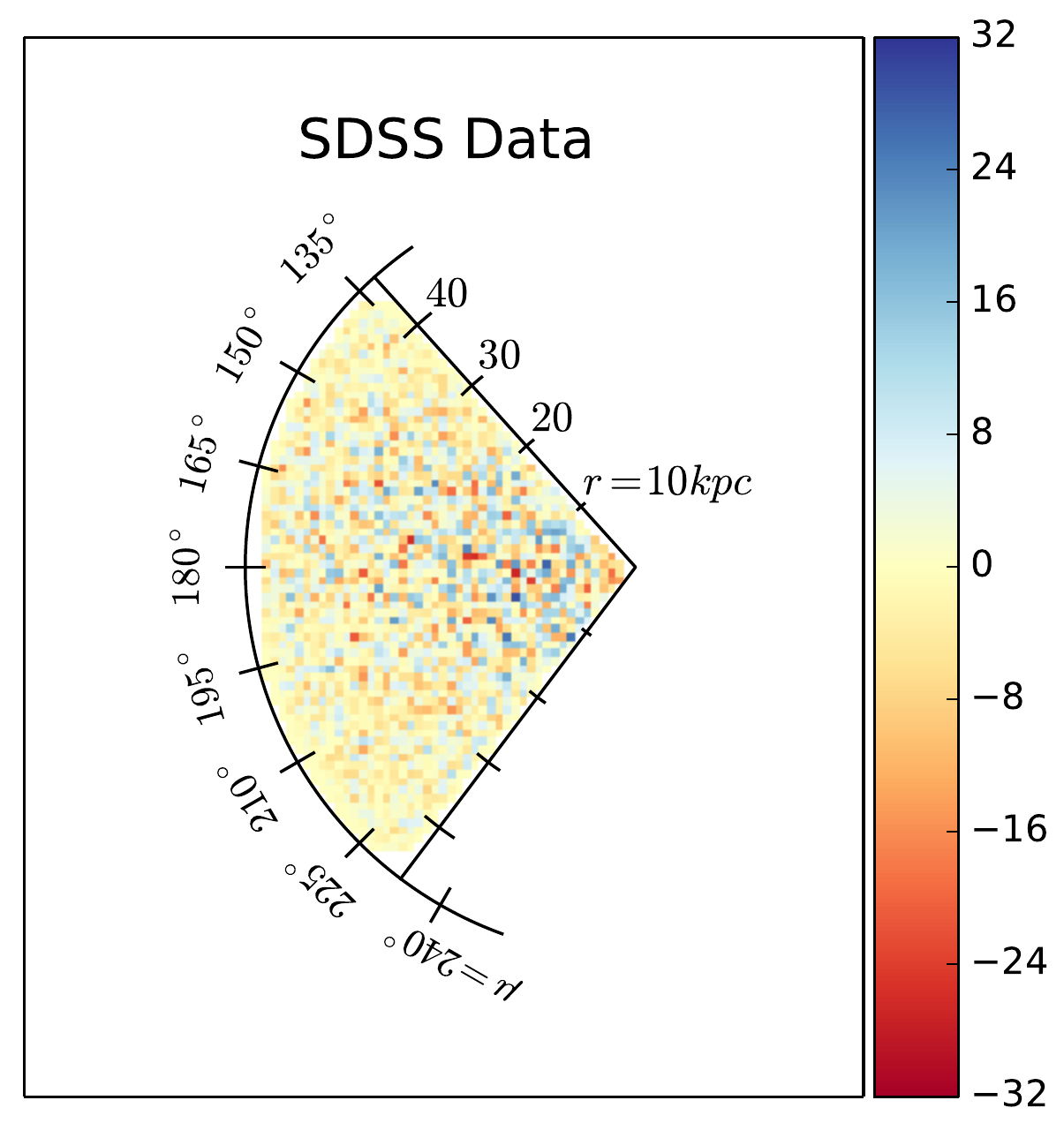}
\caption{Residual density between simulated stripe 19 and the re-simulated stripe 19.  The color represents the residual density of MSTO stars per cubic kpc in a 1 kpc by 1 kpc by $2.5^{\circ}$ pixel in the flattened (all stars are added together in the $\nu$ direction, so the volume of the stripe increases with radius from the Sun), face-on stripe.  This residual is found by subtracting the combined model in Row 3 of Figure \ref{Sim19Figure} from the combined model in Row 1 of Figure \ref{Sim19Figure}.  The residual shows that these models are very similar.}
\label{Sim19ResidualFigure}
\end{figure}

\begin{figure}
\centering
\includegraphics[width=0.33\textwidth]{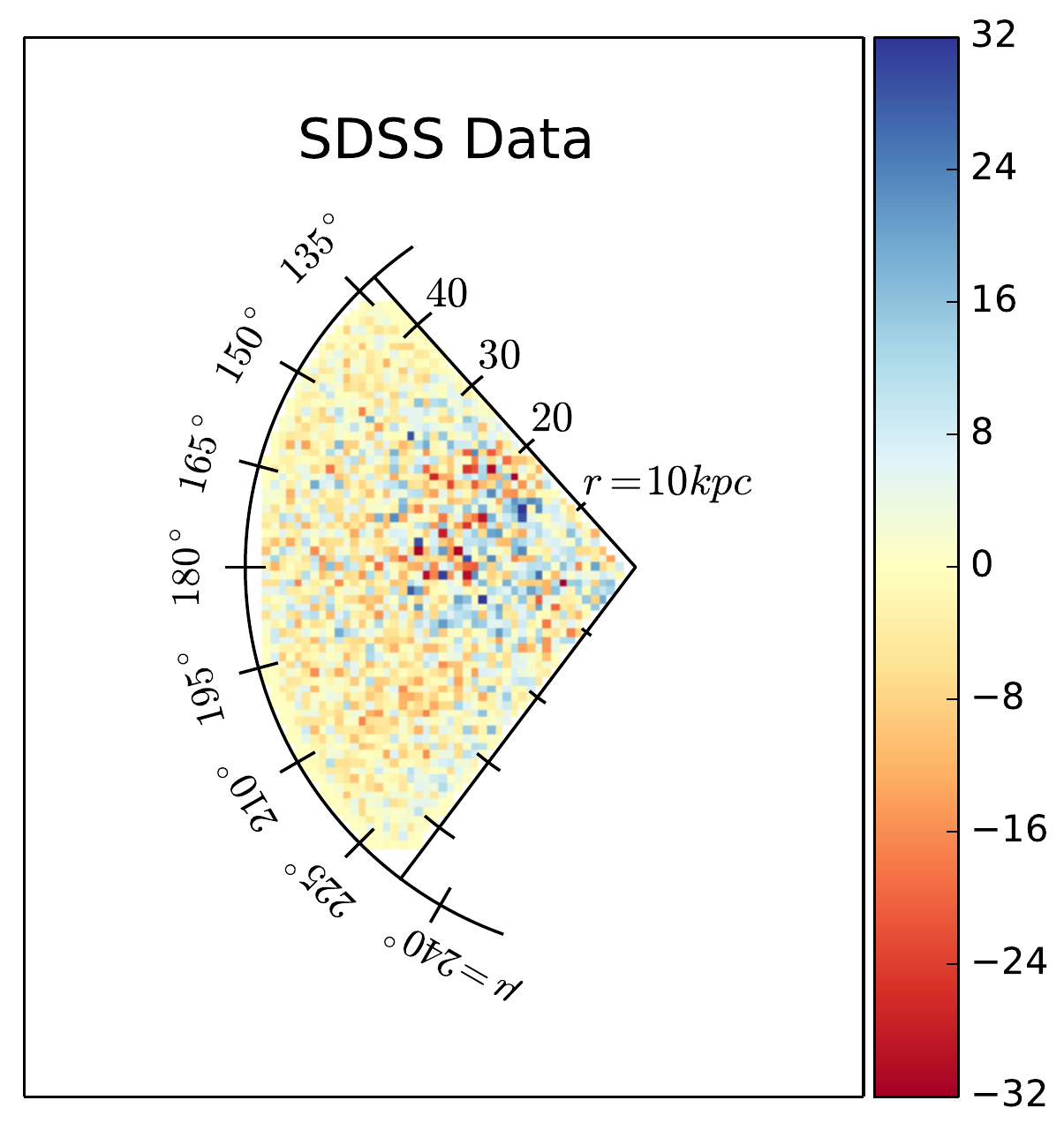}
\caption{Residual density between simulated stripe 19 with BPL background and the re-simulated stripe 19 with a BPL background.  The color represents the residual density of MSTO stars per cubic kpc in a 1 kpc by 1 kpc by $2.5^{\circ}$ pixel in the flattened (all stars are added together in the $\nu$ direction, so the volume of the stripe increases with radius from the Sun), face-on stripe.  This residual is found by subtracting the combined model in Row 3 of Figure \ref{Sim19BPLFigure} from the combined model in Row 1 of Figure \ref{Sim19BPLFigure}.  The residual shows these models differ in two primary places.  There is a blue stripe around 10 kpc followed by a red stripe around 20 kpc at $135^{\circ} < \mu < 180^{\circ}$.   The other section that shows a residual is between 25 kpc and 45 kpc and 180 degrees and 225 degrees which is dominated by red.  This section is red dominated due to the corresponding region in the BPL model which has a large deficit of stars.}
\label{Sim19ResidualBPLFigure}
\end{figure}

In this paper, we show the result of an updated version of the maximum likelihood technique developed by \cite{cole2008} and \cite{newby2013}, using updated MSTO absolute magnitude distributions from \cite{newby2011} and then fit both simulated data and real data from the SDSS.  The major conclusions are:
\begin{enumerate}
\item Our new model returns parameters that are consistent with those used to create simulations, and is ready to use on real SDSS data.  Clean separations of all of the simulated substructure in a simulated SDSS stripe 19 can be seen in Figures \ref{Stripe19Figure} and \ref{Sim19Figure}, and the comparisons of simulated to optimized parameter values can be found in Table \ref{ResultsTable}.

\item With the inclusion of the absolute magnitude distribution and selection efficiency from \cite{newby2011} we fit the Sgr stream, the trailing tail of Sgr in the north, and a third stream, which is consistent with at least some measurements of the Virgo Overdensity.  The fit density parameters for these structures in SDSS stripe 19 can be found in Table \ref{ResultsTable2}; the parameters from which the densities can be more easily derived using Equations \ref{PrelimBGDensity} and \ref{PrelimStreamDensity} are listed in Table \ref{ResultsTable3}.  The large width could have implications for the shape of the dark matter halo.

\item The presumed Sgr trailing tidal tail in north Galactic cap is much wider than the leading Sgr tidal stream, which has implications for the shape of the halo.  This stream could possibly be what is seen as the bifurcation of the Sgr tidal stream in the \cite{Belokurov2006} ``Field of Streams" image. The trailing tidal tail has a width of $\sigma = 17.6$ kpc versus $1.0$ kpc for the Sgr leading tidal tail, in our preliminary SDSS stripe 19 data fits. The distance ($48$ kpc) to the trailing tidal tail is much father than we initially expected for the ``bifurcated" stream, which has previously been reported to be $32$ kpc in \cite{Newberg2007}, $27$ kpc in \cite{Belokurov2006}, and $17$ kpc in \cite{Hernitschek2017}.  If there is a separate ``bifurcated" stream at the distance of the Sgr leading tidal tail, it was not fit by MilkyWay@home as one of the three streams.  Fitting four or more streams will be attempted in future work.
\end{enumerate}

In summary, we have developed and tested an improved algorithm for fitting stellar substructure in the Milky Way halo and demonstrated it on the data available from SDSS stripe 19 as well as on simulated test data.  We provide preliminary density information for the substructure fit in SDSS stripe 19.  We demonstrate our ability to correct for biases introduced when streams are near the limiting magnitude of the survey, and to separate overlapping substructures from each other.

Looking forward, this algorithm will be run on all of the available SDSS data in the north and south Galactic caps.  In the north there are 24 stripes of data and in the South there are 5 stripes of data, each $2.5^{\circ}$ wide.  Each of these stripes will be run through MilkyWay@home, through this effort, we will use SDSS turnoff stars to map the shape and density of the Milky Way stellar spheroid, including its major substructures.

\acknowledgments

We would like to thank the more than 200,000 MilkyWay@home volunteers for their countless hours of donated computing time, their constant community involvement with debugging and development, and their financial support over the past decade.  We would also like to thank and acknowledge contributions made by The Marvin Clan, Babette Josephs, Manit Limlamai, and the 2015 Crowd Funding Campaign to Support Milky Way Research.  This publication is based on work supported by the National Science Foundation under grant No. AST 16-15688 and the NASA/NY Space Grant fellowship.

We use data from the Sloan Digital Sky Survey.  Funding for the SDSS and SDSS-II has been provided by the Alfred P. Sloan Foundation, the Participating Institutions, the National Science Foundation, the U.S. Department of Energy, the National Aeronautics and Space Administration, the Japanese Monbukagakusho, the Max Planck Society, and the Higher Education Funding Council for England. The SDSS Web Site is \url{http://www.sdss.org/}.

The SDSS is managed by the Astrophysical Research Consortium for the Participating Institutions. The Participating Institutions are the American Museum of Natural History, Astrophysical Institute Potsdam, University of Basel, University of Cambridge, Case Western Reserve University, University of Chicago, Drexel University, Fermilab, the Institute for Advanced Study, the Japan Participation Group, Johns Hopkins University, the Joint Institute for Nuclear Astrophysics, the Kavli Institute for Particle Astrophysics and Cosmology, the Korean Scientist Group, the Chinese Academy of Sciences (LAMOST), Los Alamos National Laboratory, the Max-Planck-Institute for Astronomy (MPIA), the Max-Planck-Institute for Astrophysics (MPA), New Mexico State University, Ohio State University, University of Pittsburgh, University of Portsmouth, Princeton University, the United States Naval Observatory, and the University of Washington.  

\newpage
\bibliographystyle{aasjournal}
\bibliography{references.bib}
\end{document}